\documentclass[12pt]{amsart}
\usepackage{a4wide,palatino,epsfig,latexsym,natbib,amsmath,amsthm,graphicx,
hyperref,enumerate}

\newtheorem{remark}{Remark}
\newtheorem{example}{Example}

\newtheorem{result}{}

\newcommand{\R}{{\mathbb R}}

\newcommand{\Z}{{\mathbb Z}}

\begin{document}

\title{Polytopes, graphs and fitness landscapes}  

\begin{abstract}
Darwinian evolution can be illustrated
as an uphill walk in a landscape,
where the surface consists of genotypes,
the height coordinates represent
fitness, and each step corresponds to a point
mutation.
Epistasis, roughly defined as the dependence between
the fitness effects of mutations,
is a key concept in the theory
of adaptation. Important recent approaches
depend on graphs and polytopes.
Fitness graphs are useful for describing
coarse properties of a landscape,
such as mutational trajectories and the number of peaks.
The graphs have been used for relating
global and local properties of fitness
landscapes.
The geometric theory of
gene interaction, or the shape theory, is
the most fine-scaled approach
to epistasis.
Shapes, defined as triangulations
of polytopes for any number of
loci, replace
the well established concepts
of positive and negative
epistasis for two mutations.
From the shape one
can identify
the fittest populations,
i.e., populations where
allele shuffling (recombination)
will not increase the mean fitness.
Shapes and graphs provide complementary
information. The approaches make
no structural assumptions about the
underlying fitness landscapes,
which make them well suited for
empirical work.
\end{abstract}

\author{Kristina Crona}  


\maketitle


\section{Introduction}
The fitness landscape
was originally intended
as a simple metaphor
for an intuitive understanding
of adaptation \citep{w}.
Adaptation 
can be pictured
as an uphill
walk in the fitness landscape,
where height represents
fitness and where
each step is between
similar genotypes.
The concept 
of a fitness landscape
has been formalized in somewhat
different ways \citep{bpse} and the current theory 
is extensive. Kaufman's NK model \citep{kw},
block models \citep{mp, o2006},
as well as
random (rugged or uncorrelated) fitness landscapes
\citep{k,kl,fl,rbj,pk}
have been especially
influential in biology.
Early work in the field was
primarily motivated by theoretical
considerations, such as
the relation between global and
local properties of fitness landscapes.
However, it may not be clear
if the classical models
apply in a particular 
empirical context. 
The underlying
assumptions, such
as a block structure
of the fitness landscape,
may or may not hold.

Some recent approaches
do not make any 
structural assumptions
about the fitness 
landscapes. We will consider 
the geometric 
theory of gene interactions
and fitness graphs.
We define fitness
as the logarithm of the
expected reproductive success.
There are different definitions
of fitness in the literature 
\citep{moh}. Epistasis means
that fitness is not linear.
For instance, the 
combination of two beneficial mutations
may result in a double mutant with much
higher fitness, as compared to a linear 
expectation from the fitness of the 
wild-type, and the two single mutants. 
Such positive epistasis
is common for drug resistance
mutations, for example
antibiotic resistance
mutations \citep[e.g.][]{gmc}. 
It is not difficult
to analyze the two-loci case,
but it is less obvious
how to quantify, classify
and interpret epistasis 
for several loci.

The most fine-scaled approach to
gene interactions is the
recently developed geometric theory \citep{bps}.
The theory extends the usual concept of epistasis
for two mutations to any number of loci in
the strict sense that all gene interactions 
are reflected. The shapes,
as defined in the geometric theory,
has the role of positive and negative
epistasis for two mutations. 

In contrast to the sensitive 
shapes, a fitness graph is determined 
by the fitness ranks of the genotypes only.
Qualitative information such as if 
"good+good=better" or 
"good+good= not good"
for two single mutations
are reflected by the fitness graphs.
From the graphs one can immediately understand 
the coarse properties of the landscapes, including
the number of peaks. We argue that both 
the geometric theory and fitness graphs
are well suited for empirical work.
Moreover, to some extent shapes and fitness graphs
provide complementary information.
Shapes are relevant for recombination and
fitness graphs for mutational trajectories.

In many real populations at most two 
alternative alleles occur at each locus, or  
a biallelic assumption is a reasonable simplification.
Throughout the chapter, we will 
consider biallelic $L$-loci populations.
Let $\Sigma=\{0,1\}$ and let $\Sigma^L$
denote bit strings of length $L$.
$\Sigma^L$ represents the genotype space.
In particular,
\[
\Sigma^2=\{ 00, 01, 10, 11 \}
\text{ and }
\Sigma^3=\{ 000, 001, 010, 011, 100, 101, 110, 111 \}.
\]
The {\emph{zero-string}} denotes the string with 
zero in all $L$ positions, and
the {\emph{1-string}} denotes the string with 1 in all $L$ 
positions. We define a {\emph{fitness landscape}} as
a function $w:\Sigma^L\mapsto \mathbb{R}$,
which assigns a fitness value to
each genotype. The fitness
of the genotype $g$ is denoted $w_g$.
The metric we consider is the Hamming distance,
meaning that the distance between two genotypes equals
the number of positions where the genotypes differ. 
In particular, two genotypes are adjacent,
or {\emph{mutational neighbors}}, if
they differ at exactly one position.

A walk in the
fitness landscape corresponds
to a Darwinian process in a
precise way. Consider a monomorphic
population, i.e., a population
where all individuals
have the same genotype,
after a recent change in the environment.
Such a genotype is a called
the {\emph{wild-type}}. Assume
that in the new environment
the wild-type no longer has optimal fitness.
Under the assumption of the strong-selection weak-mutation regime
(SSWM), a beneficial mutation will go to fixation
in the population before the next mutation occurs.
It follows that the population is 
monomorphic for most of the time.
The adaptation process can be described as
a sequence of genotypes, all of which
became fixed in the population at some point in time.

For instance, let $00$ denote the
wild-type, assume that the single mutants
$10$ an $01$ have higher fitness
than the wild-type, and that
the double mutant $11$ has the highest fitness of the
four genotypes.
The two possible adaptation
scenarios for a population are
\[
00 \mapsto 10 \mapsto 11 \text{ and }
00 \mapsto 01 \mapsto 11.
\]
Each scenario corresponds to an uphill walk,
which ends at the genotype $11$.
The example illustrates that we can think of a
Darwinian process as a walk in the fitness landscape,
where each step represents a beneficial mutation going
to fixation in the population. Adaptation is
not deterministic, but
fitness has to increase by each step.
The described model of adaptation
has been widely used and relies on approaches
developed in \citet{g1983,g1984,ms}.

The chapter is structured as follows.
The topic for Section 2-5 is fitness graphs,
where most results depend on \citet{cgb}.
The topic for Section 6-10 is the geometric theory of gene interactions,
where most results depend on \citet{bps},
and triangulations of polytopes \citep{drs}.
Section 11 compares fitness graphs and shapes, as defined in the geometric 
theory. Section 12 is a discussion. 
For more background,
including proofs, we refer to
\citet{cgb, bps, drs}.

\section{Fitness graphs and sign epistasis}
With reference to the landscape
metaphor, an {\emph{adaptive step}} in 
the fitness landscape corresponds to a change 
in exactly one position of a string so that the 
fitness increases strictly. 
An {\emph{adaptive walk}} is a sequence of adaptive steps. 
A {\emph{peak}} in the fitness landscape has the property 
that there are no adaptive
steps away from it, i.e., a genotype is at a peak if all 
mutational neighbors have lower fitness as compared to the genotype.
The following concepts are 
central as well, in particular they are
useful for relating the number 
of peaks to local observations.

For $L \geq 2$, given a string and two
positions, exactly four strings can be obtained
which coincide with the original string except    
(at most) at the two positions. 
Denote such a set of four strings
\[
ab, Ab, aB, AB,
\]
according to the two positions of interest, and
assume that $w_{ab}$ is minimal.
{\emph{Sign epistasis}} means that
\[
w_{AB} < w_{Ab} {\mbox{ or }} w_{AB} < w_{aB}.
\]
{\emph{Reciprocal sign epistasis}}
interactions means that
\[
w_{AB} < w_{Ab} {\mbox{ and }} w_{AB} < w_{aB}.
\]
Fig. 1 shows the four possibilities
under our assumption that $w_{ab}$ is minimal.
\begin{figure}
\begin{center}
\includegraphics[scale=0.7]{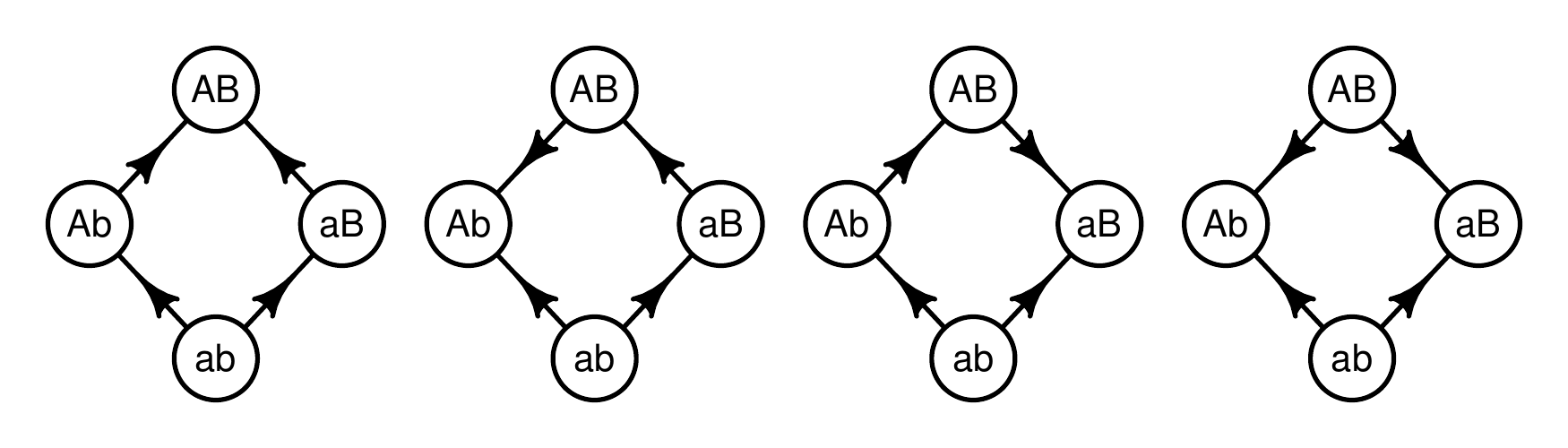}
\end{center}
\caption{The arrows point toward the more fit genotype. The graphs
represent no sign epistasis, two cases with sign epistasis
but not reciprocal sign epistasis, and one case with
reciprocal sign epistasis.}
\end{figure}

Sign epistasis is by no means 
rare for microbes according to several 
studies 
\citep[e.g.][]{djk, wwc, wdd, bes, fkd, ssf, gmc}.
In particular, sign epistasis
occurs for antibiotic resistance mutations,
as well as for HIV and malaria.
In fact, existing studies suggest that
{\emph{absence}} of sign epistasis
is exceptional for systems associated
with drug resistance for $L\geq 4$. 

Sign epistasis is
of clinical importance for several reasons. 
A recent approach for preventing 
and managing resistance problems
takes advantage of both sign epistasis and 
variable selective environments \citep{gmc}.
Another aspect of managing drug resistance
is to find constraints for orders in which 
mutations accumulate from
genotype data \citep{djk, bes}. 
A constraint could
be that a particular mutation is 
selected for (meaning that that the mutation
is beneficial) only if a different mutation has already 
occurred.
The existence of constraints implies sign epistasis.
Indeed, if a particular
mutation is beneficial regardless of background,
then it can occur before or after other mutations. 
Moreover, sign epistasis is relevant
for predictions of how populations
will adapt \citep{wdd}.

Fitness graphs are useful for
the empirical problems mentioned, 
as well as for more theoretical 
problems, including the relation
between global and local properties
of fitness landscapes (see Section 3). 
If one can order a set
of genotypes by decreasing
fitness, one has determined
the {\emph{fitness ranks}}. More
fine scaled information, such as
relative fitness values, may 
not be known.
A fitness graph 
compares the fitness ranks of 
mutational neighbors.  
For simplicity, whenever we
use fitness graphs we assume that $w_s \neq w_{s'}$
for any two strings  $s$ and $s'$ which
differ in one position only.

Roughly, consider the zero-string 
as the starting point (possibly the wild-type), 
and each non-zero position of a string as
an event, i.e., that a mutation has occurred.
Under these assumptions the fitness graph
coincides with the Hasse-diagram of the 
power set of events, except that each 
edge in the Hasse-diagram
is replaced with an arrow toward 
the string with greater fitness.

For a formal definition,
a fitness graph is a directed
graph where each node corresponds
to a string of $\Sigma^L$.
The fitness graphs has $L+1$ levels.
Each string such that
$\sum s_i=l$ corresponds to a node
on level $l$ in the fitness graph.
In particular, the
node representing the
zero-string is at the
bottom, the nodes representing
strings with exactly one non-zero
position, including
$
10 \cdots 0,
$
are one level above,
the nodes representing
strings with exactly two non-zero
positions, including
$
110 \cdots 0,
$
are on the next level,
and the 1-string is at the top.
Moreover, the nodes are ordered
from left to right
according to the lexicographic order
where $1>0$ of the corresponding
strings (see e.g. Fig. 5).
A directed edge
connects each pair of nodes such that
the corresponding strings differ
in exactly one position.
The edge is directed toward
the node representing the
more fit of the two genotypes. 

\begin{remark}
Unless otherwise stated,
the words "level", "up", "down"
"above" and "below" refer to
fitness graphs.
In particular,
notice that a higher level does not
imply greater fitness.
\end{remark}

For $L\geq 2 $, given a string and two positions,
consider the four strings which coincide 
with the original string 
except in (at most) the two positions.
We call the strings a {\emph{type 2}}
system if there is reciprocal sign epistasis,
a {\emph{type 1}} system if there is sign epistasis,
but not reciprocal sign epistasis,
and a {\emph{type 0}} system if there is 
no sign epistasis.

For interpretations of
general fitness graphs,
it may be helpful to
first analyze
the two-loci case
shape in some detail. There exist
exactly 14 fitness graphs
for biallelic two-loci systems
(see Fig. 2),
where 4 are type 0
systems, 8 type 1 systems, and
2 type 2
systems.
One verifies the following result.

\begin{remark}
For two-loci, type 0, 1, and 2 systems have the
following properties:
\begin{enumerate}
\item
A type 0 system can be
rotated so that all arrows
point up.
\item
A type 1 system
differs from a cycle by exactly
one arrow.
\item A type 2 system
have two nodes such that
all edges are directed
toward them, and two nodes
such that no edges are
directed toward them.
\end{enumerate}
\end{remark}

\begin{figure}
\begin{center}
\includegraphics[scale=0.7]{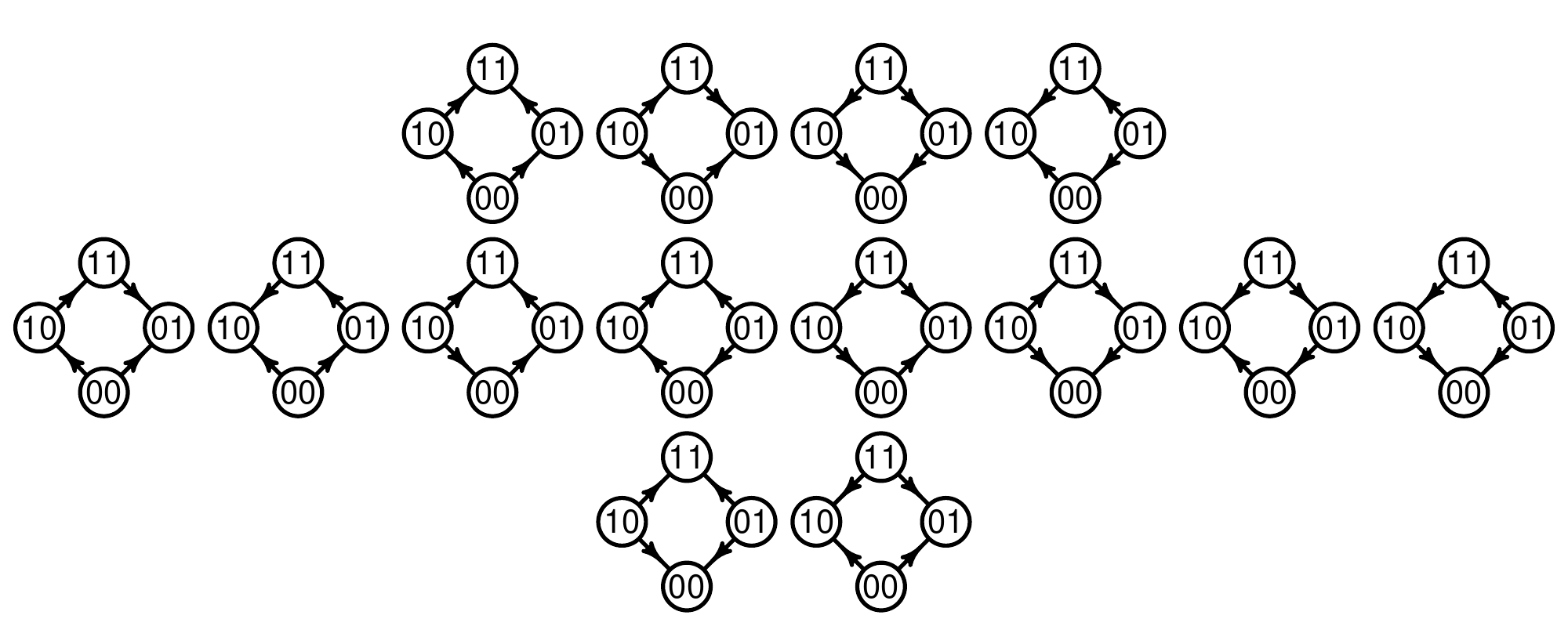}
\end{center}
\caption{
For a fitness graph, the arrows point
toward the genotype of
greater fitness. There exist exactly 14 fitness graphs
for biallelic two-loci systems,
where the type 0 systems are on the
first row, the type 1 systems on the
second row, and the type 2 systems
on the third row.}
\end{figure}

The observations from the two-loci
case should make it easy to identify
type 0, 1 and 2 systems for general
fitness graphs. Fig. 3 and 4 show
fitness graph for 3-loci systems.
Fig. 3a has type
0 systems only, Fig. 3b type
0 and 2 systems,
Fig. 4a type 0, 1 and 2 systems,
and Fig. 4b type 2 systems only.
Fig. 5 shows a fitness graph for
a 4-loci population, where there
are several type 2 systems,
including 0001, 0101, 0011, 0111.

\begin{figure}
\begin{center}
\includegraphics[scale=0.4]{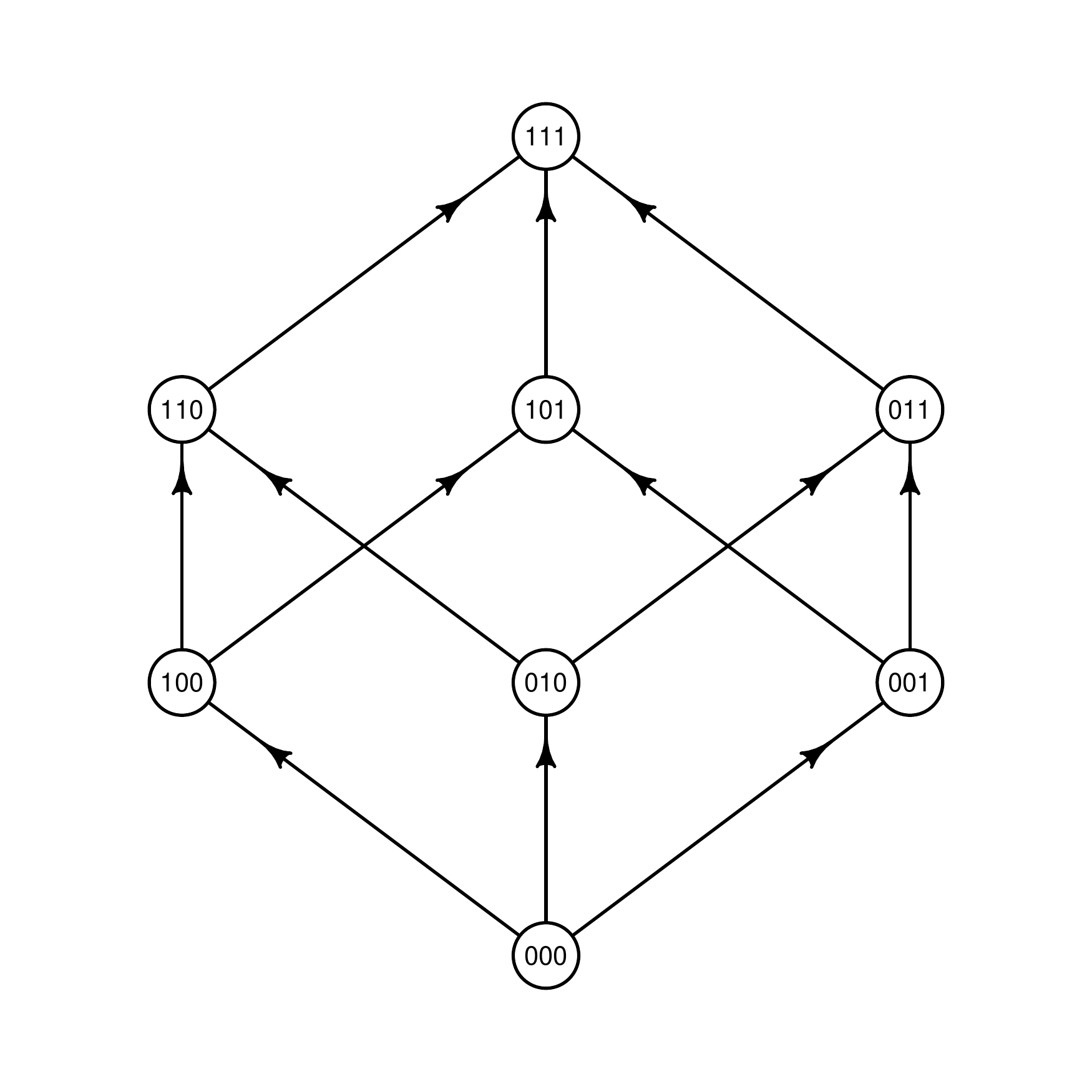}
\includegraphics[scale=0.4]{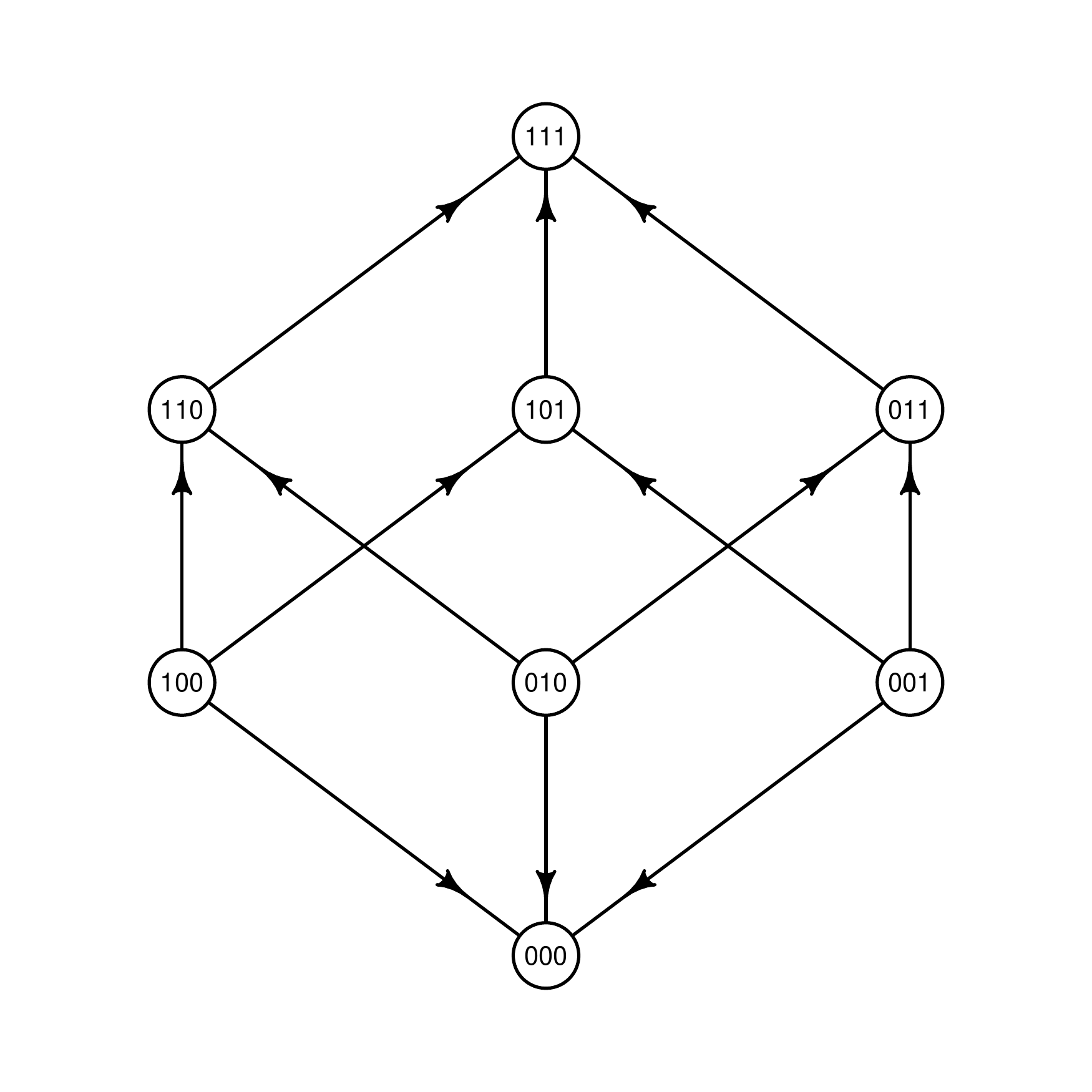} \\
\end{center}
\caption{  
A fitness graph shows
sign epistasis and the peaks.
The graph in Fig. 3a has type 0 systems only.
The graph in Fig. 3b has type 0 and type 2 
systems, but no type 1 systems.
}
\end{figure}

\begin{figure}
\begin{center}
\includegraphics[scale=0.4]{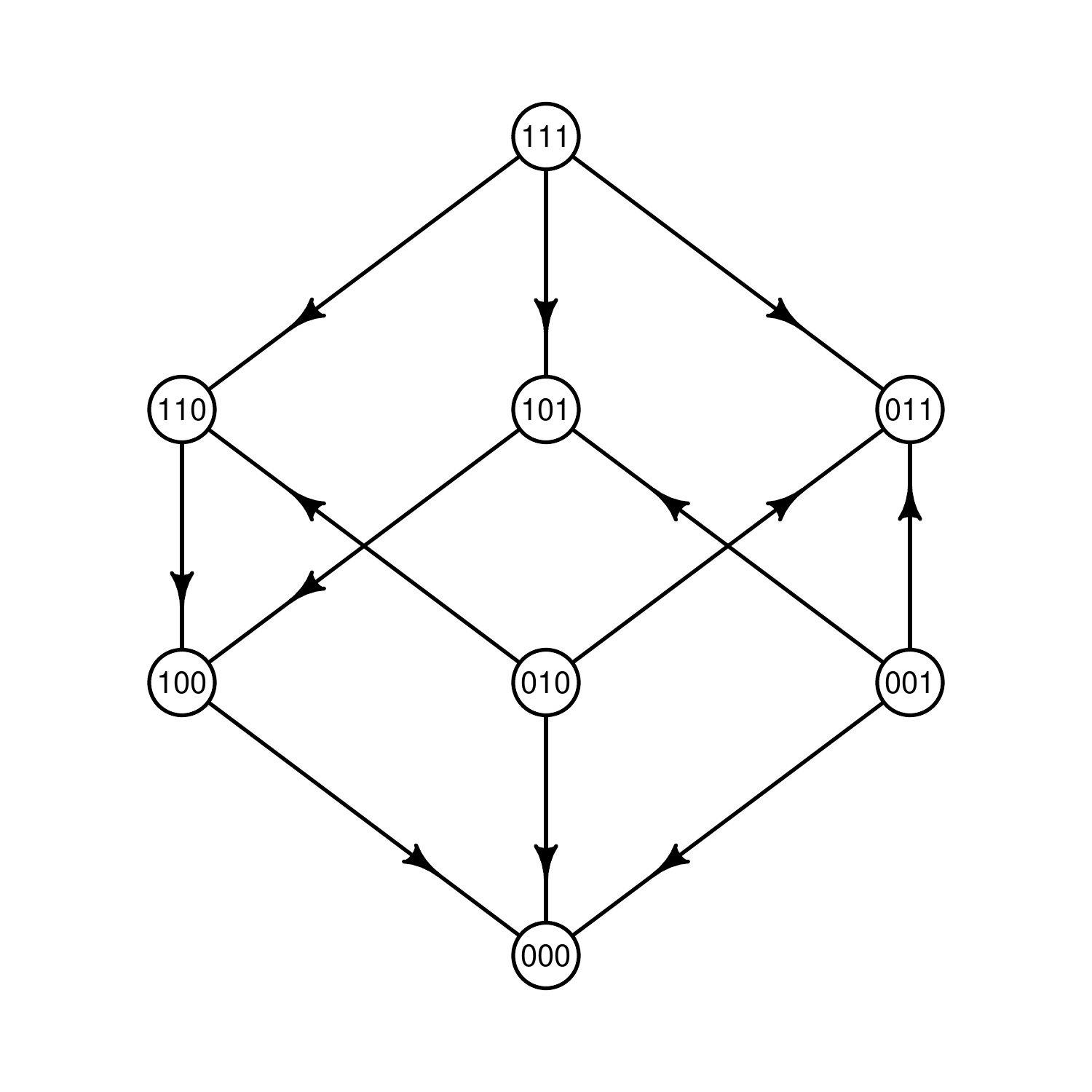}
\includegraphics[scale=0.3]{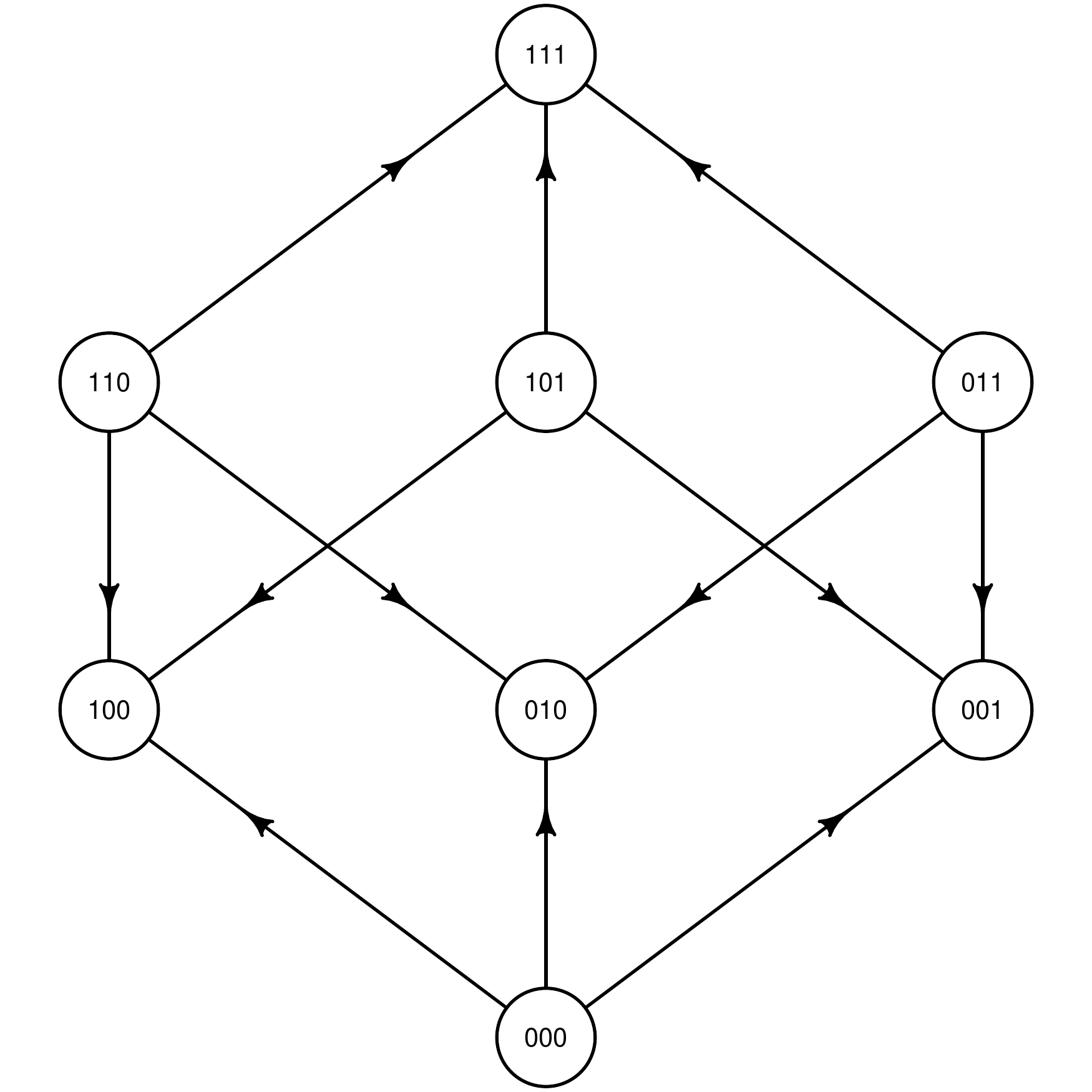}
\end{center}
\caption{
The graph in Fig. 4a has type 0, type 1 and
type 2 systems. The graph in Fig. 4b has
type 2 systems only, and the corresponding
fitness landscape has four peaks.}
\end{figure}

\begin{figure}
\begin{center}
\includegraphics[scale=0.4]{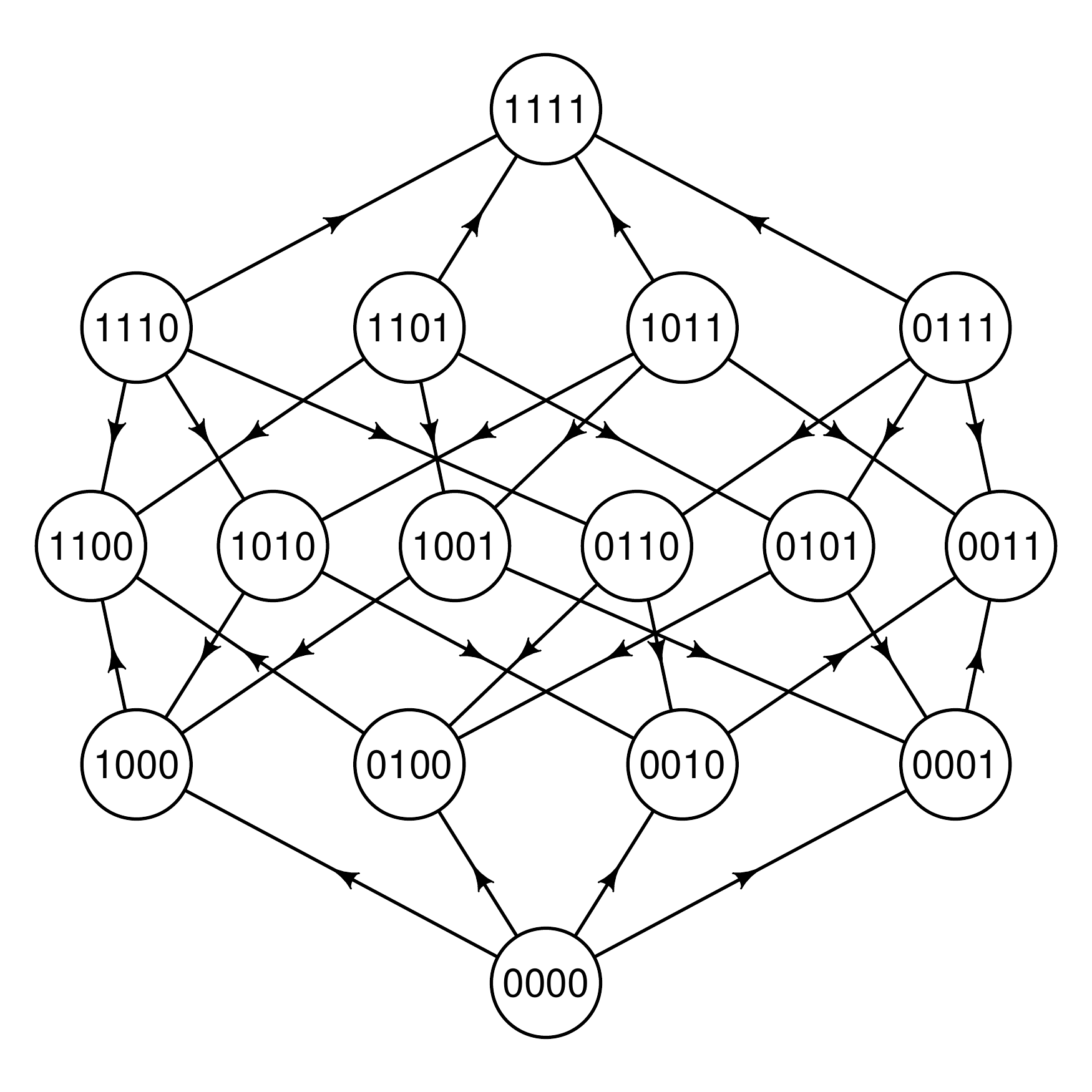}
\end{center}
\caption{The fitness landscape
has peaks at $1100$, $0011$ and $1111$,
whereas all triple mutants (mutants on the third level)
have low fitness}
\end{figure}

\section{Fitness graphs and theoretical results}
Fitness graphs have mostly been used in empirical
work \citep[e.g.][]{dpk, fkd, ssf, gmc}.
However, we will indicate how they can
be used in theoretical arguments, and mention some 
results where the proofs depend on fitness graphs.

It is known that one can have
$2^{L-1}$ peaks in a fitness landscape \citep[e.g.][]{h} 
and this number is an upper bound.
The proof is elementary, 
and we will not give the details.
However, we will construct
fitness landscapes with
the maximal number of
peaks using fitness graphs.

\begin{example}
For any $L$,
consider the fitness
graph where the
edges are directed
up from level 0
to 1, down from level 1 to 2,
up from level 2 to 3, and so on.
The fitness graph
in Fig. 4b
is an example.
Notice that the graph
corresponds to fitness landscapes
with 4 peaks, i.e., 
the maximal number
of peaks for $L=3$. 
In general,
all nodes at level
$1, 3, 5 \dots $ 
are at peaks,
and such fitness
graphs correspond
to fitness landscapes
with  exactly $2^{L-1}$
peaks.
\end{example}

Recent work relates
global and local properties
of fitness landscapes \citep{pkw, psk, cgb}.
This topic is of interest,
since most empirical studies of fitness 
landscapes concern local properties,
including sign epistasis.
It has been shown
that multipeaked
fitness landscapes have
type 2 systems \citep{psk}.
The converse is not true.
However, a sufficient condition for
multiple peaks can be phrased
in terms of type 1 and 2 systems.
More precisely, the following result 
was proved using fitness
graphs.

\begin{result}{\bf{(Crona et al., 2013)}}
If a fitness landscape has type 2 systems and
no type 1 systems, then it has multiple peaks.
\end{result}
It follows that the landscapes
corresponding to
Fig. 3b and 4b have multiple peaks. 

Fitness graphs are efficient
for analyzing mutational trajectories.
We will state a result regarding
accessible mutational trajectories
from \cite{wwc}. A brief proof of 
the result using
fitness graphs was given
in \citet{cgb}, but the original 
proof does not use fitness graphs.  

We call the global maximum of the landscape 
"the fitness peak".
Moreover, define a {\emph{general step}} similar to
"adaptive step", except that the fitness may
decrease. A {\emph{general walk}},
as opposed to an "adaptive walk" is
a sequence of general steps.
If a general walk between two nodes has minimal
length, we call it a {\emph{shortest walk}}.

\begin{result} {\bf{(Weinreich et al., 2005)}}
\begin{enumerate}
\item
The following conditions
are equivalent for a fitness
landscape.
\begin{enumerate}
\item[(i)]
Each general step toward the fitness peak, i.e.,
a step that decreases the graph theoretical distance to the peak,
is an adaptive step.
\item[(ii)]
Each shortest general walk to the fitness peak is
an adaptive walk.
\item[(iii)]
The fitness landscape has no type 1 or 2 systems.
\end{enumerate}
\item If the equivalent conditions in (1) are
satisfied, then each adaptive walk to the fitness peak is a
shortest general walk.
\end{enumerate}
\end{result}

A fitness landscape satisfying the equivalent conditions (i)--(iii)
above is referred to as {\emph{a fitness landscape lacking genetic
constraints on accessible mutational trajectories}}
in \cite{wwc}. 
For $L=3$, the fitness graph in
Fig. 3a corresponds to this category
of landscapes.
Fitness landscapes lacking genetic
constraints on accessible mutational trajectories
can be represented by
fitness graphs where all arrows
are up.
For brevity, we will refer to
"all arrows up landscapes".

It is important to notice that the concept
of an all arrows up landscape
is biologically meaningful. 
Even if a landscape is single peaked,
type 1 systems may cause the adaptation process
to be slower since not all shortest general walks to the peak
are adaptive walks. However, for all arrows up
landscapes, there are no local obstacles for the 
adaptation process.

\section{Fitness graphs and recombination}
Recombination can generate
new genotypes in a population.
Under some circumstances, recombination
will speed up adaptation. 
An early hypothesis about
the possible advantage of recombination
concerned double mutants
of high fitness, where the
corresponding single mutations are
deleterious. It was suggested that 
recombination could
generate such  
double mutants.
In terms of fitness 
graphs this case 
can be described as
a type 2 system, where
the wild-type is
at a fitness peak.    
However, the hypothesis was
immediately criticized,
and described as a
"widespread fallacy"
by Muller \citep{crow}. 
The two single mutations
being deleterious,
it seems unlikely that 
the the corresponding
genotypes
would appear and recombine
to the double mutant. The 
(current) consensus
is that under most circumstances
recombination will not
be of any use in the situation
described, i.e., 
for a two-loci type-2 system,
where the wild-type is at a 
peak (see also \citet{lop}). 
However, using fitness
graphs we will
argue that
recombination
could be an advantage
in somewhat related 
cases where $L\geq 3$.

The topic of recombination
is involved with subtle 
differences between effects on 
the population level and the gene 
level. For instance, it is 
theoretically possible that 
recombination is 
beneficial for a population
and at the same time 
recombination suppressors 
could be selected for
(see e.g. \citet{ol} for comments
and references). 
We do not intend to develop
new theory, or describe
existing knowledge of recombination
in any detail. For an overview of the 
field, we refer to \citet{ol}.
Our goal is to point out mechanisms 
specific for $L\geq 3$ loci which 
should be considered for an analysis 
of the effect of recombination. This 
is justified since the field is 
dominated by work in the two-loci case, 
or mechanisms which can be reduced to
the two-loci case.

It has been suggested that
recombination has an especially
strong impact in 
{\emph{structured populations}} 
\citep[see e.g.][]{mol}. A populations 
is structured, as opposed to well mixed,
if the genotype frequencies
varies between geographic locations.
In particular, if a population 
is subdivided into local
subpopulations with
some migration between them,
then recombination could be
advantageous.

We will sketch a model within
this framework, which we call a 
{\emph{puddles and flood
population}}. We mainly have
microbes in mind, for example 
bacteria. Assume that the local 
subpopulations live in puddles, 
and the subpopulations
are homogeneous for 
most of the time.
Occasionally, there is 
a flood where 
the contents of 
the local puddles
get thoroughly mixed. 
After a flood,
life proceeds
as usual in the puddles
for an extended period, until
the next flood.
Under these assumptions,
genotypically different 
subpopulations 
are likely to mix,
so that recombination
can generate new genotypes.

\begin{example}
Consider the fitness graph
in Fig. 5, and assume that
0000 is the wild-type.
Both  1100 and 0011 are at peaks,
whereas the triple mutants
are less fit as compared to adjacent
double mutants. 
For a puddles and flood
population, recombination
of double mutants may result
in 1111. 
In this case 
recombination could
speed up adaptation.

Notice
that in the absence of
recombination,
one could obtain
1111 from 1100,
only if there is  
a double mutation,
since the triple
mutants are not fit.
\end{example}

\begin{example}
Consider 
the fitness graph in
Fig. 4b.  Assume that
000 is the wild-type.
From the fitness graph,
the single mutants
100, 010, 001 are
at peaks. Under the
assumption that 111
has maximal fitness,
recombination could
speed up adaptation.
However, two recombination
events are necessary.
For instance,
recombination of
100 and 010 could
result in 110.
Then recombination of 
110 and 001
could result in 111.
\end{example}
Notice that there is an
important difference
between Examples 2 and 3.
For instance, consider the
outcome for a puddles and
flood population
where no more than
two puddles mix
at the time.
Then one could 
obtain 1111 by
recombination in Example 2.
Indeed, if an 1100
population and an 0011
population mix, 
recombination could
produce the 1111 genotype. 

In contrast, consider 
Example 3 under the same
restriction (no
more than two puddles mix).
If say
an 100 and 010 population
mix, one could obtain
110 by recombination.
However, 110 
is selected against
so that 111 
is unlikely to appear
under most circumstances.
(On the other hand, if
{\emph{several}} 
puddles mix, 
one could get
a mixture of 
the single mutants
100, 010 and 001, and 
recombination
could result in 111).

Consider all arrows up fitness
landscapes where the 1-string
has maximal fitness. 
Then one could obtain the 1-string
from a sequence of single mutations.
However, for a puddles and flood
population, recombination could
speed up adaptation. This is
because the
process of accumulating $L$
single mutations could be 
time consuming.

\begin{example}
For an all arrows up
$L$-loci fitness landscape
where considerably
more than $L$ 
puddles tend to mix during
a flood period,  
one could  
obtain the 1-string
already after
one flood period.
\end{example}

The examples
described
are theoretical 
constructions.
It is not obvious if
Examples 2
and 3, or similar cases,
occur frequently
enough in nature
for having much of an
impact.
A first question to ask
for a population,
is how frequently it
happens that
"good+good=not good"
for single mutations.
This type of problems is the
topic for the next section.

\section{Fitness graphs and other qualitative measures}
In order to determine if one has a 
reasonable chance to find fitness graphs
of the types described in the previous sections,
the following qualitative concept 
\citep{cgb} may be useful.

We define $B$ and $B_p$ as follows.  
{\emph{The set $B_p$ consists of all
double mutants such that both corresponding
single mutations are beneficial.}}

{\emph{The set $B \subseteq B_p$
consists of all double mutants
in $B_p$ which are more fit
than at least one of the
corresponding single mutants.}}

The {\emph{qualitative measure of  additivity} 
for a fitness landscape is the ratio $\frac{|B|}{|B_p|}$.
Notice that $\frac{|B|}{|B_p|}=1$
for all arrows up landscapes.

Fitness landscapes are defined
as {\emph{additive}} if 
fitness effects of mutations sum.
For example,
if 
\[
w_{00}=1, w_{10}=1.2, w_{01}=1.3,
\]
then additive fitness implies
that $w_{11}=1.5$ (since $0.5=0.2+0.3$,
so that the fitness effects of two
mutations sum). 
Notice that fitness
is additive exactly if
\[
w_{11}-w_{10}-w_{01}+w_{00}=0 \,.
\]
By definition, fitness is additive 
exactly if there is no epistasis.
One may consider all arrows up landscapes
as the qualitative correspondence
to additive fitness landscapes.

Antibiotic resistance landscapes 
for a particular 4-loci system 
and 9 selective environments 
were studied in \citet{gmc}.
More precisely, all combination
of the TEM-1 mutations
L21F, R164S, T265M and E240K
were considered. The length
of TEM-1 is 287, i.e.,
TEM-1 can be represented as a sequence of
287 letters in the 20-letter
alphabet corresponding
to the amino acids.
The notation 
L21F means that the amino acid 
Leucine (L) at position 21 has 
been replaced by the amino acid Phenylalanine (F).
The mutations R164S, T265M and E240K
are defined similarly, using
the standard notation for amino acids.
The mean value of $\frac{|B|}{|B_p|}$ 
for the 9 selective environments was 0.57.

In contexts where
the relative fitness values of genotypes
are not known, qualitative
concepts can still be used.
It is valuable to understand qualitative information
for several reasons.
Fitness ranks tend to be easier to determine as compared to 
relative fitness, and from records of
mutations one can sometimes draw 
conclusions about fitness ranks without
making measurements. We argue that much can be learned
about fitness landscape from existing records of mutations,
in particular from drug resistance mutations.
However, one needs to be able to interpret qualitative
information, a theme developed in \citet{cps}
with applications to antibiotic resistance,
see also \citet{cgb}.

The qualitative measure of additivity 
is coarse. If relative fitness values 
can be determined, one may want to consider 
quantitative fitness differences as well. 
For a quantitative measure of 
additivity, we refer to the concept 
"roughness" \citep{ch, aih}. Briefly,
additive fitness landscapes have 
roughness 0, and any deviation from 
additivity implies that 
the roughness is greater than 0.  

\section{Shapes}
Throughout the remainder of the
chapter, we consider
biallelic $L$-loci populations,
where we assume that 
all $2^L$ genotypes occur in the populations 
(for a comment regarding this simplification,
see Section 10),
and a fitness landscape 
\[
w:\Sigma^L\mapsto \mathbb{R},  \, \Sigma=\{0,1\}.
\]
Moreover, for $L=2 \text{ and }3$ 
we use the following orders of the
genotypes (from left to right): 
\[
00, 01, 10, 11
\text{ and  }
000, 001, 010, 011, 100, 101, 110, 111.
\]

Most empirical studies of epistasis
for several loci focus on 
the average curvature (Fig. 6)
or pairwise gene interactions
using ANOVA methods \citep{bps}.
For beneficial mutations,
{\emph{antagonistic epistasis}}
means that the combined effect  
of mutations are less than the sum of individual 
effects,
whereas  {\emph{synergistic epistasis}}
means that the combined effect of
mutations exceeds the sum
of individual effects.
It has been claimed that
antagonistic epistasis
dominates for beneficial mutations
in nature \citep[e.g.][]{kdp}. 
Antagonistic and synergistic epistasis
are defined analogously
for deleterious mutations (Fig. 6).
A motivation for the interest 
in average effects of mutations
is the connection to recombination.
According to standard
models, antagonistic epistasis
for beneficial mutations
sometimes implies an advantage
for recombination \citep{ol}.
\begin{figure}
\begin{center}
\includegraphics[scale=0.2]{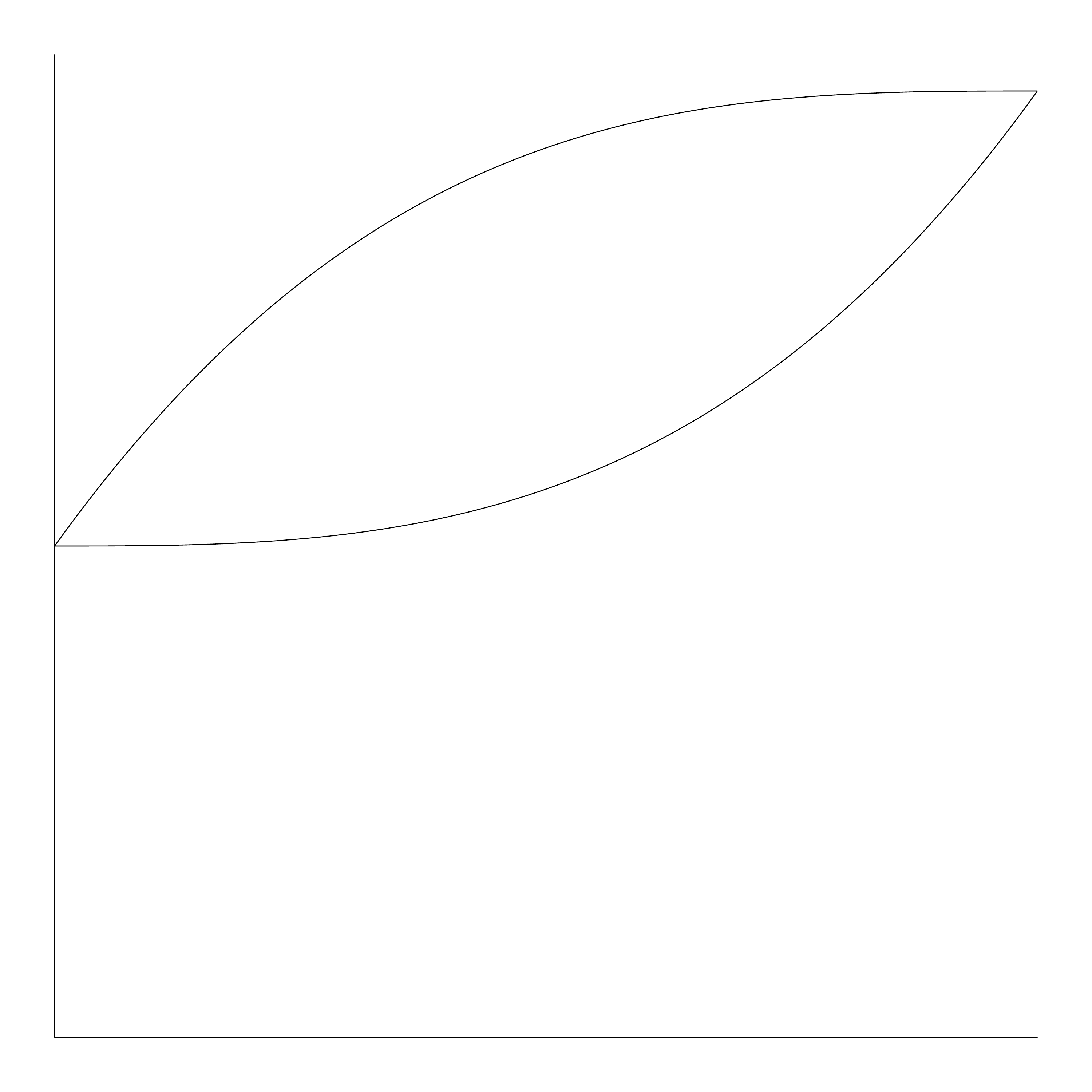}
\includegraphics[scale=0.2]{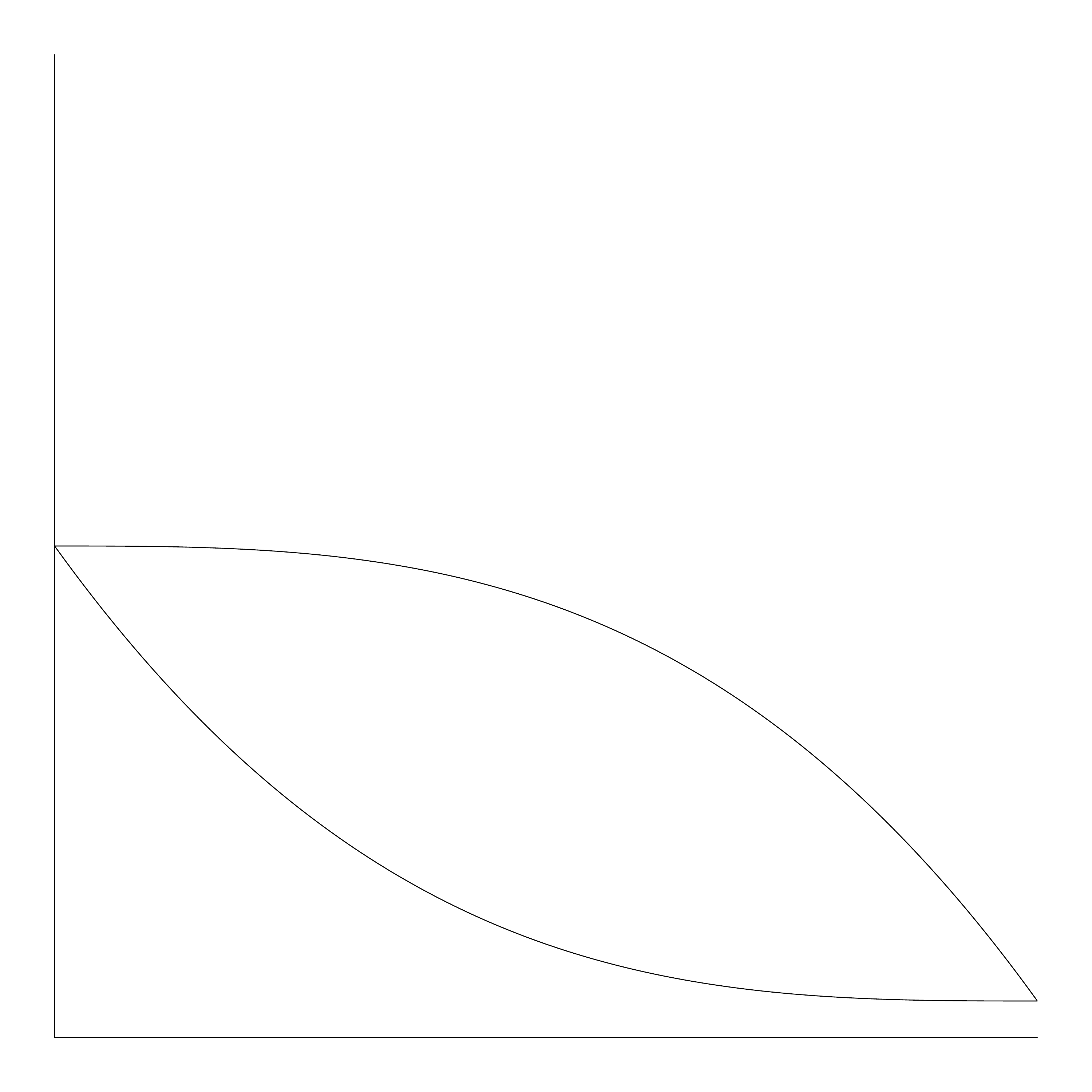}\\
\end{center}
\caption{
The number of mutations increases along the
horizontal axis, and the fitness increases along
the vertical axis. For Fig. 6a,
the mutations are beneficial.
The upper curve corresponds to antagonistic epistasis,
and the lower curve to synergistic epistasis.
For Fig. 6b, the mutations are deleterious.
The upper curve corresponds to synergistic epistasis
and the lower curve to antagonistic epistasis.}
\end{figure}

Conventional summary statistics
for epistasis have their limitations. The average 
curvature may obscure a diversity of 
interaction types, and pairwise tests fail to 
discover curvature at genetic distances 
greater than two.
The most fine-scaled approach
to gene interactions is the
geometric theory, introduced in \citet{bps}. 
The theory reveals 
all the gene interactions, and it
depends on triangulations of
polytopes. For mathematical
background we refer to \cite{drs},
see also \cite{z} for the
general theory about
polytopes. The geometric approach has
revealed previously unappreciated
gene interactions for HIV, {\emph{Escherichia-coli}}
and in some other cases \citep{bps, bpse},
and the approach is relevant for recombination.

We will start with an informal introduction to the
geometric theory, where the main purpose is to 
provide an intuitive understanding 
and some geometric interpretations. More formal 
descriptions are given in the next sections.

Roughly, a triangulation of a polygon
is a subdivision of the polygon into triangles.
A triangulation of the $L$-cube
is a subdivision of the cube into simplices
(triangles if $L=2$, tetrahedra if 
$L=3$, pentachora for $L=4$, and so on).
We will use some concepts which
are defined in terms of populations.
If one groups individuals
into classes of identical genotypes,
a population can be described as the
frequencies of the genotypes.
The {\emph{fitness of a population}}
is defined as the
average fitness of all individuals.

First consider the case $L=2$. 
Let
\[
\Delta=\{ (p_{00}, p_{01}, p_{10}, p_{11})\in [0,1]^4
\, : \, p_{00} + p_{01} + p_{10} + p_{11}=1\}
\]
denote the population simplex. A population 
is given as a point in $\Delta$.
The {\emph{genotope}} for $L=2$
is the square with vertices $00, 01, 10, 11$.
We denote this genotope $[0,1]^2$,
and interpret a point $v=(v_1,v_2) \in [0,1]^2$
as the allele frequencies of the population,
where $v_1$ denotes the frequency of 1's at the first locus,
and $v_2$ the frequency of 1's at the second locus.

For a simple example,
consider a population where half of the individuals 
have genotype 00, and the other half 11. Then the allele 
frequency vector is $(0.5, 0.5)$.
For a population where half of 
the individuals
have the genotype 01 and
the other half 10, the allele frequency
vector is  $(0.5, 0.5)$ as well. However, 
the average fitness may differ between the two 
populations.

We will analyze examples in more detail.
\begin{example}
Consider $v=(0.4, 0.8) \in [0,1]^2$
and the populations $p^1=(0.2, 0.4, 0, 0.4) \in \Delta$
and  $p^2=(0, 0.6, 0.2, 0.2) \in \Delta$.
One verifies that both populations have the
allele frequencies described by $v$;
indeed, adding the contributions of 1's for 
$p^1$ and the first locus
gives  $0+0.4=0.4$, and for the second locus
$0.4+0.4=0.8$. The contributions 
for $p^2$ gives $0.2+0.2=0.4$ for the first 
locus, and $0.6+0.2 = 0.8$ for the second.
\end{example}

Let $\rho$ denote a corresponding map from 
the population simplex
$\Delta$ to the genotope $[0,1]^2$,
where
\[
\rho(p_{00}, p_{01}, p_{10}, p_{11})=
\left( p_{10}+ p_{11}, p_{01}+ p_{11} \right).
\]
Then $\rho$ maps a point of the population
simplex to the allele frequencies, 
where $p_{10}+p_{11}$ equals
the frequency of 1's at the first locus,
and $p_{01}+ p_{11}$ equals the frequency
of 1's at the second locus.
Notice that $\rho(p^1)=\rho(p^2)=v$ in 
the previous example.

Given a fitness landscape and
a vector $v\in [0,1]^2$,
a {\emph{fittest population}}  
$p\in \Delta$ has maximal 
fitness among populations 
such that the allele frequencies 
are described by $v$. Moreover,
$p$ is unique for $L=2$,
except in the case when fitness is additive
(see also the comment after Case 2 below).
For a fittest population,
one cannot increase the fitness
by shuffling around alleles.
The biological significance is
immediate, since such allele
shuffling relates to recombination.

For a geometric interpretation,
the fitness landscapes $w$ (usually) induces
a triangulation of the genotope $[0,1]^2$.
This triangulation is the {\emph{shape}} of the fitness landscape.
The critical property of the triangulation
is that for  $v\in [0,1]^2$,
the genotypes that occur in the fittest population
are the vertices of the triangle which contains
$v$. The corresponding result holds 
for any $L$.
We will first describe
the triangulations, and then
give a geometric interpretation
of shapes. As remarked,
fitness is additive exactly if
\[
w_{11}-w_{10}-w_{01}+w_{00}=0 \,.
\]
For simplicity, we call the case
where 11 has higher fitness as compared
to a linear expectations
positive epistasis, 
and similarly for negative epistasis.

{\bf{ Case 1: }}(positive epistasis) If
\[
w_{11}-w_{10}-w_{01}+w_{00}>0,
\]
then the triangulation induced by the
fitness landscape has
$00-11$ diagonal, meaning
that the triangles are $\{00,01,11\}$
and $\{00,10,11\}$ (Fig. 7).

{\bf{ Case 2: }}(negative epistasis) 
If
\[
w_{11}-w_{10}-w_{01}+w_{00}<0,
\]
then the induced triangulation of the genotope   
has $10-01$ diagonal meaning
that the triangles are $\{00,01,10\}$ and $\{01,10,11\}$
(Fig. 7).

For positive epistasis, $p^1$ in Example 5
is a fittest population. Indeed, 
positive epistasis
implies that whenever one replaces 10 and 01 genotypes by
00 and 11 genotypes, the result is
increased average fitness of the population.
However, for $p^1$ the proportions of 
00 and 11 genotypes are maximal (in the sense that one cannot replace
10 and 01 genotypes by 00 and 11 genotypes), so that
$p^1$ has maximal fitness given the allele frequency vector
$v=(0.4, 0.8)$. 

For positive epistasis, notice that
$v=(0.4, 0.8)$ is a point in the triangle 
$\{00,01,11\}$ and that the genotypes
of $p^1$ are 00,01, 11. 
Moreover, 10 and 01 are not on the same triangle,
which indicates that one
can increase fitness by replacing them
with other genotypes. These observations 
illustrate how shapes and fittest 
populations relate for $L=2$.

For a geometric interpretation
of shapes,
consider the genotope $[0,1]^2$
and the four points above the vertices
of $[0,1]^2$, such that the height coordinates
correspond to fitness.
The four points are vertices of a tetrahedron (Fig. 7).
The upper sides of the tetrahedron (marked with 
different patterns) project onto two triangles of
$[0,1]^2$. The projections describe the
triangulation induced by $w$. The left picture 
corresponds to positive epistasis, and the right to 
negative epistasis. This construction should make sense, 
since the triangulation obtained as projections of
the upper faces of the tetrahedron has the critical 
property for all fittest populations. More precisely, for any $v\in[0,1]^2$, 
a fittest population consists of vertices of the 
triangle which contain $v$. The fitness 
landscape almost always induces a triangulation 
of the genotope  $[0,1]^2$ as described. Such a triangulation
is a generic shape. 
The exceptional (non-generic) case is when fitness is additive.

\begin{figure}
\begin{center}
\includegraphics[width=5.3in]{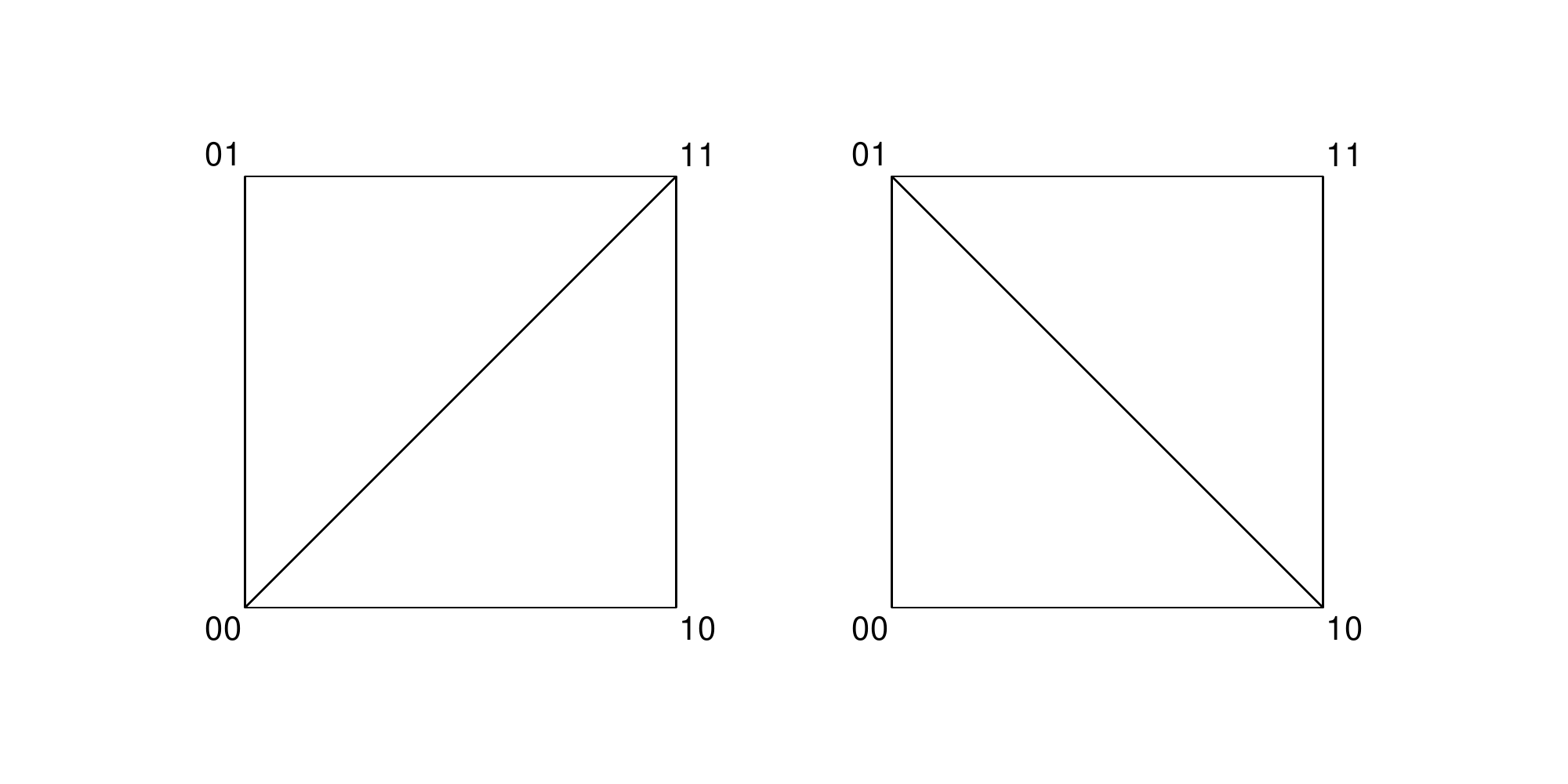}
\includegraphics[width=5.3in]{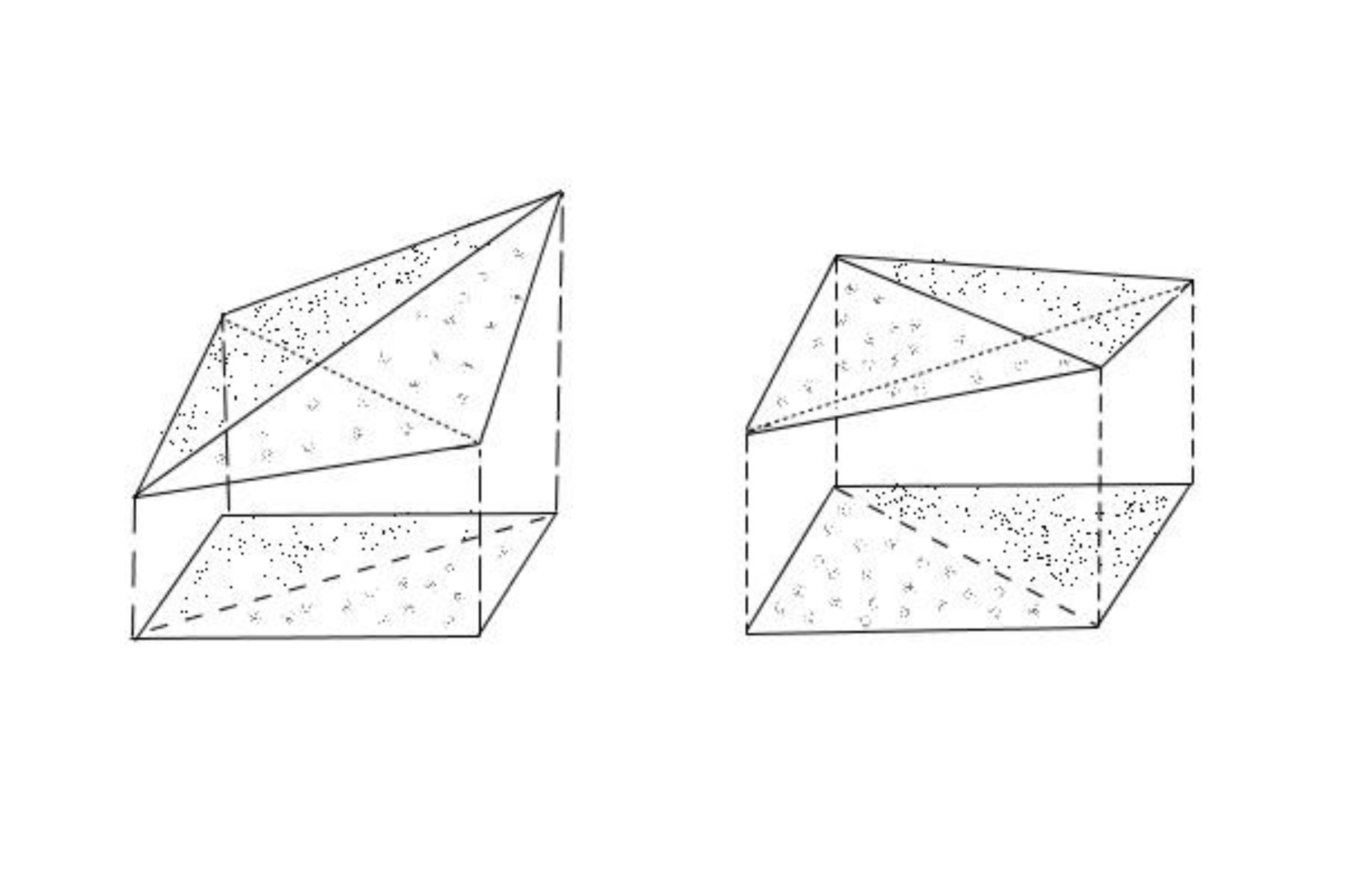}  
\end{center}
\caption{The upper pictures show the   
triangulations of the genotopes (the squares) 
in Case 1 (positive epistasis) and 2 (negative
epistasis).
The lower left picture shows
the tetrahedron above the genotope
in Case 1, where the height coordinates
correspond to the fitness of
the four genotypes under consideration.
The projections of the upper sides of the tetrahedron
describe the triangulations. The lower
right picture shows how the triangulation
is induced in Case 2.}
\end{figure}

In general,
consider a biallelic $L$-loci system
and the fitness landscape $w$.
The {\emph{genotope}} is the $L$-cube $[0,1]^L$, 
where the vertices represent the genotypes.
As in the two-loci case, 
let $\Delta$ denote the population simplex and
let $\rho$ denote the corresponding map from 
$\Delta$ to $[0,1]^L$.
For a fixed $v \in [0,1]^L$,
consider the
linear programming problem
\[
\max \, \{ \, p \cdot w : \rho(p)=v  \}.
\]
A solution gives the maximal population fitness, i.e.,
the maximum of $p \cdot w$,
given the allele frequency vector $v$ (since $\rho(p)=v)$.
Consequently, finding the fittest population
translates to solving this linear programming problem.

If we let $v$ vary,
we get the following
parametric linear programming problem
\[
\tilde{w} (v) = \max \{ \, p \cdot w : \rho (p)=(v)
{\text{ for all }} v \in [0,1]^L \}.
\]
The domains of linearity of $\tilde{w}$
do almost always constitute a triangulation 
of the genotope \citep[Chapter 2]{drs}. The shape of 
the fitness landscape is the triangulation 
of $[0,1]^L$ induced by the fitness 
landscape $w$. 
The geometric interpretation is analogous to the two-loci case,
so that the triangulation is obtained as the projections 
of the upper faces of the polytope constructed from 
the fitness landscape. Moreover, in the generic
case, i.e., if the shape of the fitness landscape
is a triangulation, the fittest population is
unique for a given allele frequency vector $v$.  
More precisely, 
the genotypes that occur in 
the fittest population are the vertices of the 
simplex which contains $v$.

For the two-loci case,
the geometric theory does
not contribute anything 
new, since there exist 
only two triangulations
corresponding to the two types
of epistasis in the usual sense.
However, for $L=3$,
there are 74 generic shapes corresponding
to triangulations of the
cube (see Section 9).

Not all triangulations can be obtained
from a fitness landscape.
A triangulation is {\emph{regular}}
if it is induced by some fitness landscape.
Fig. 8 shows a non-regular triangulation.
This is the smallest non-regular 
triangulation.
\begin{figure}
\begin{center}
\includegraphics[scale=0.2]{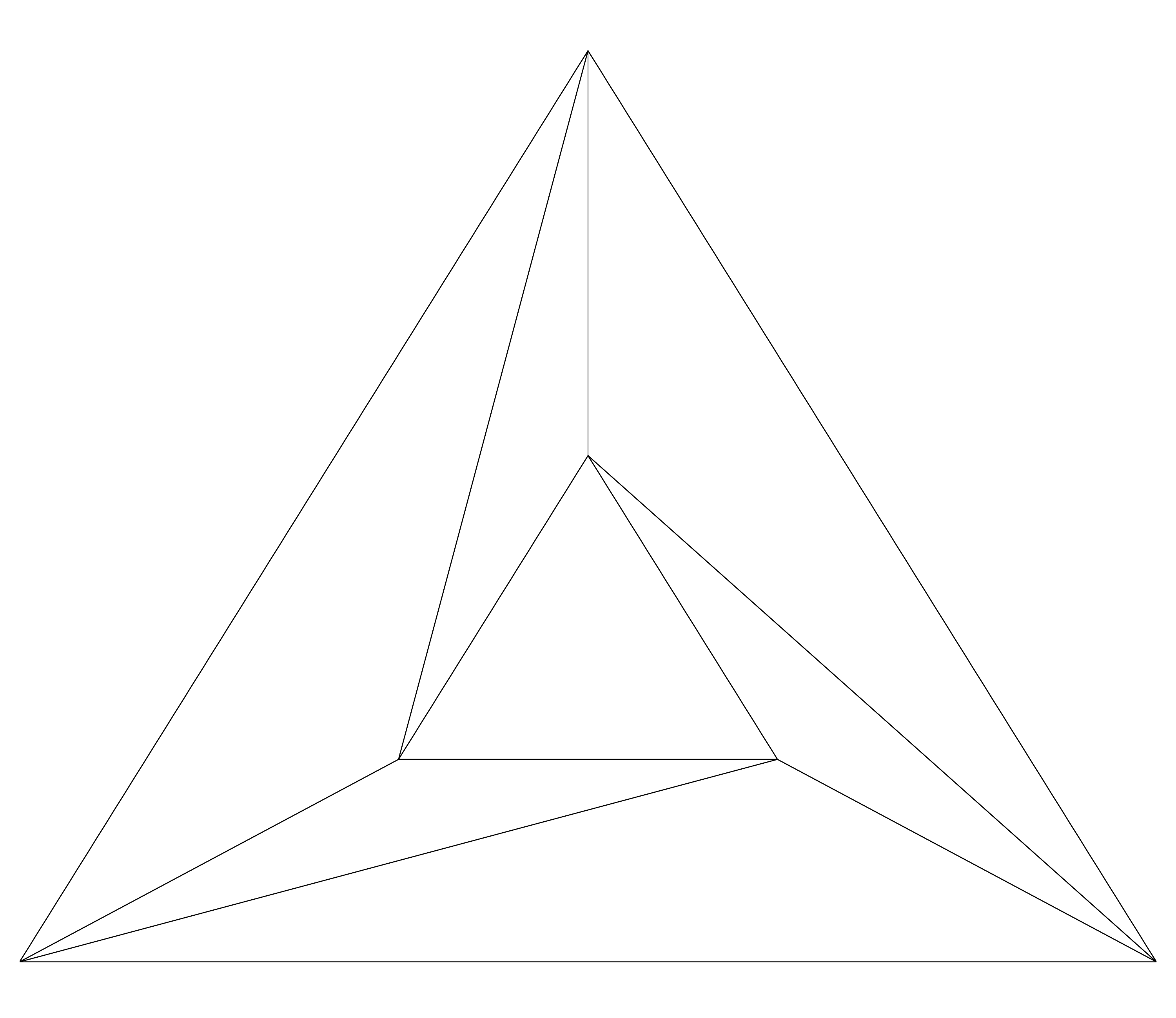}
\end{center}
\caption{A non-regular triangulation. This triangulation
cannot be induced from a fitness landscape.}
\end{figure}
In the literature, a regular triangulation is
described as a triangulation which is
induced by a cost vector. Then the linear 
programming problem concerns minimizing the cost,
and the triangulations are obtained
as projections of all lower faces
of the polytope constructed from
the cost vector. Since our topic is fitness 
landscapes, we think in terms
of maximal fitness rather than minimal costs.

\section{Shapes and flips}
For $L>2$,
there are many possible shapes.
It may seem that shapes
are difficult to apply
in empirical biology,
due to ambiguity from measurement
errors. However, the geometric theory 
comes with a structure.
Shapes may be similar or completely
different, and the relation between
shapes can be described in a systematic
way. We start with intuitive descriptions.
Briefly, a {\emph{flip}}, sometimes referred
to as a geometric bistellar flip,
is a minimal change between triangulations.
Fig. 9, 10 and 11 show flips.
For the two-loci case, the two triangulations
corresponding to positive and negative
epistasis differ by a flip.

For an overview of how all 
triangulations of a polytope
are related, one can consider the
{\emph{flip graph}}.
The nodes of the graph are
the triangulations, and edges connect
triangulations which differ by a flip.
Fig. 12. shows the flip graph of a hexagon.
The graph theoretical distance between
triangulations can be considered
a measure of how closely 
related the triangulations
are. Some caution is necessary
if one is primarily interested
in regular triangulation, since
a regular triangulation 
may be transformed into
a non-regular triangulation
by a flip.

\begin{figure}
\begin{center}
\includegraphics[scale=0.5]{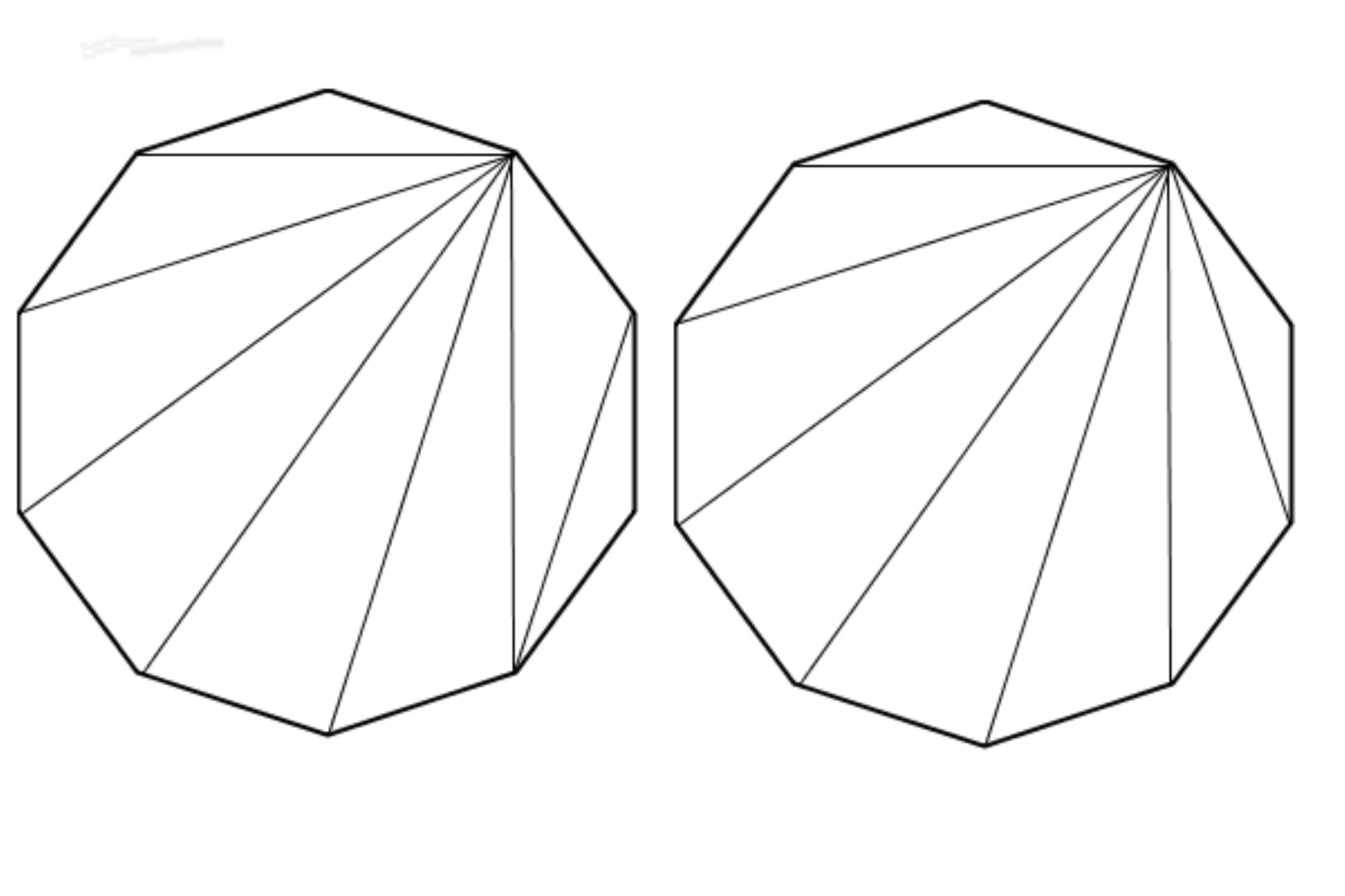}
\end{center}
\caption{The left triangulation can be transformed
into the right triangulation by a flip.}
\end{figure}

\begin{figure}
\begin{center}
\includegraphics[scale=0.4]{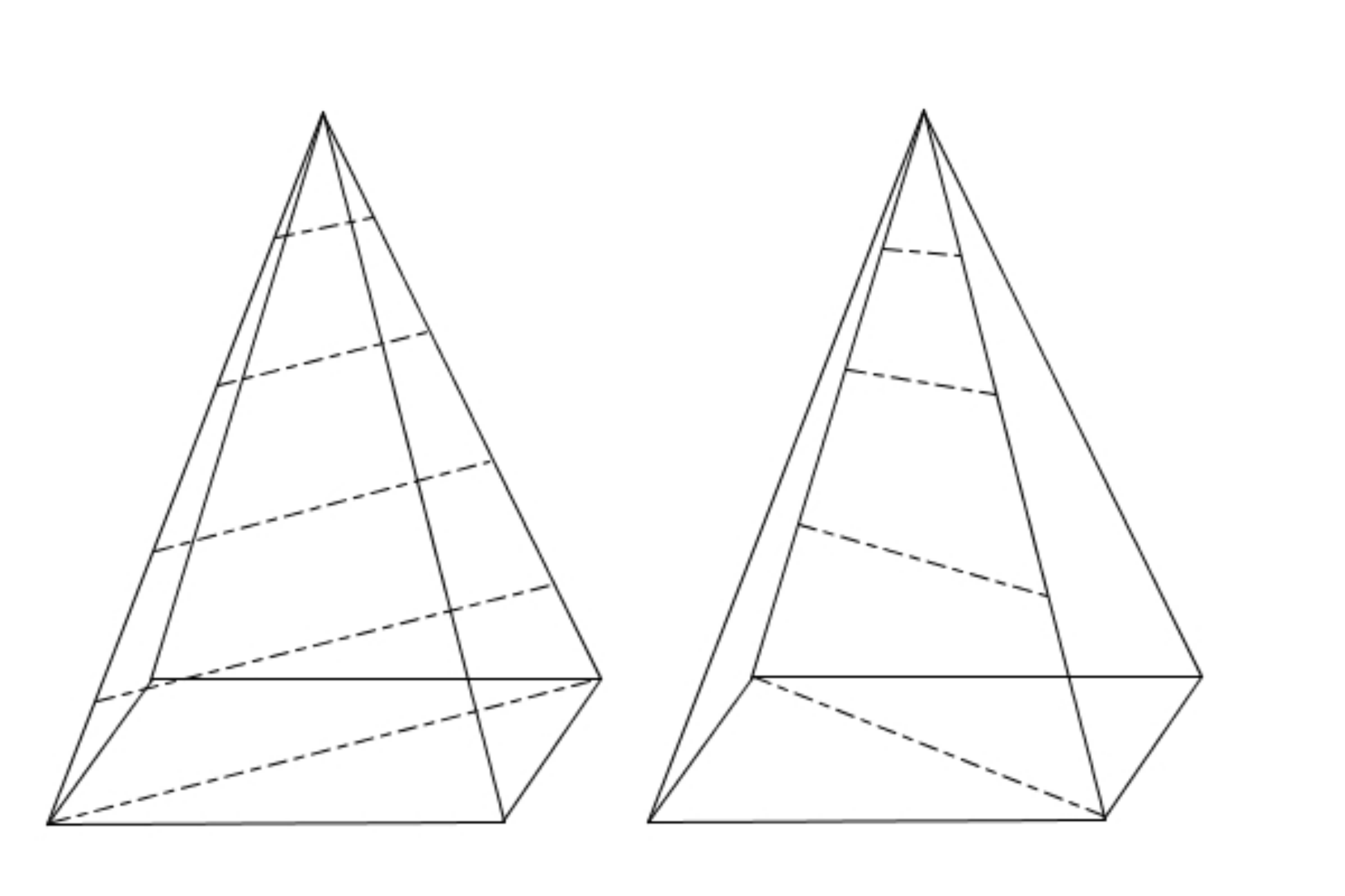}
\end{center}
\caption{The dashed lines indicate the triangulations.
The triangulations differ by a flip.}
\end{figure}

\begin{figure}
\begin{center}
\includegraphics[scale=0.5]{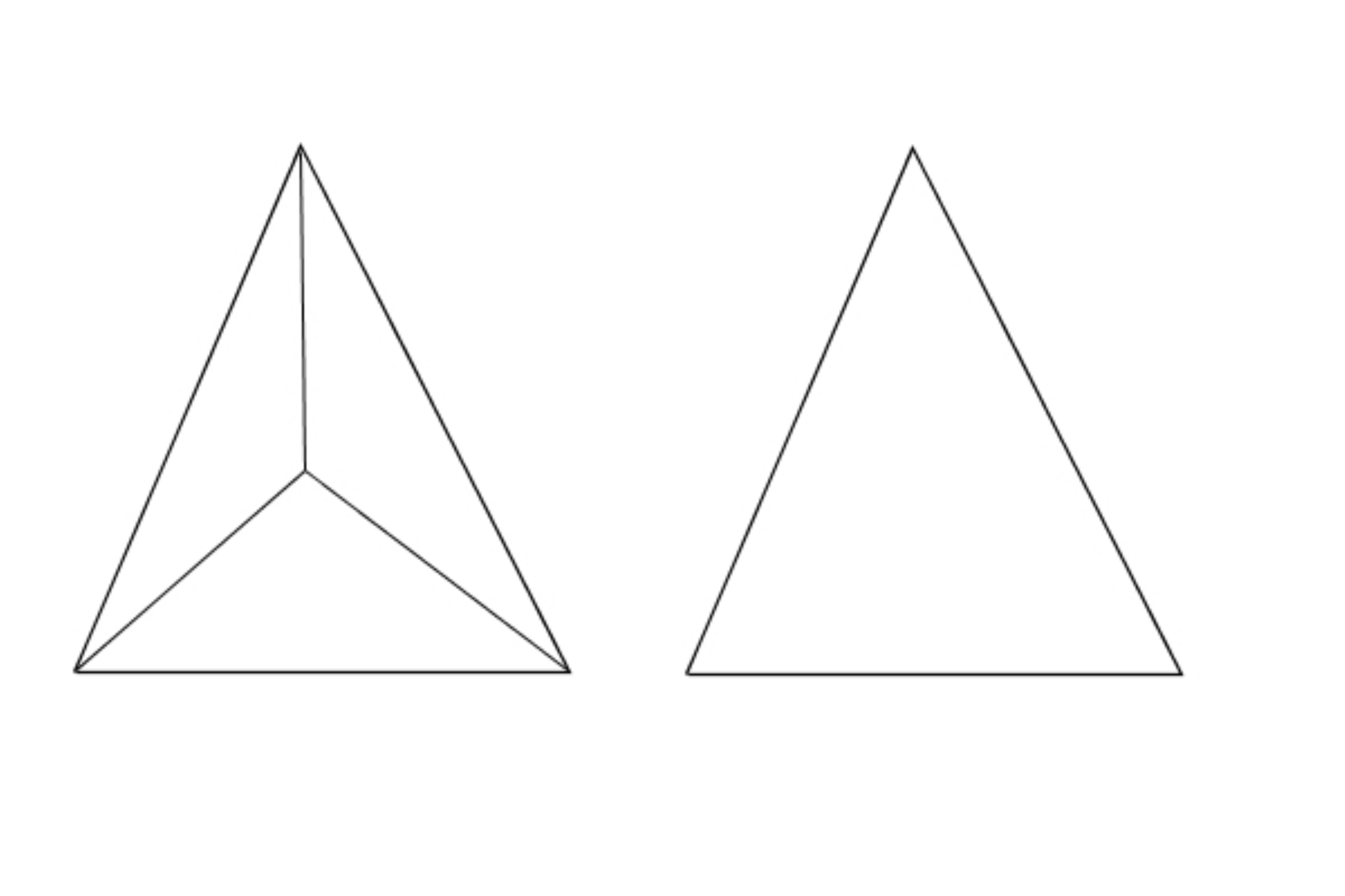}
\end{center}
\caption{
The triangulations differ by a flip,
and the number of triangles are
different for the two triangulations.}
\end{figure}

\begin{figure}
\begin{center}
\includegraphics[scale=0.5]{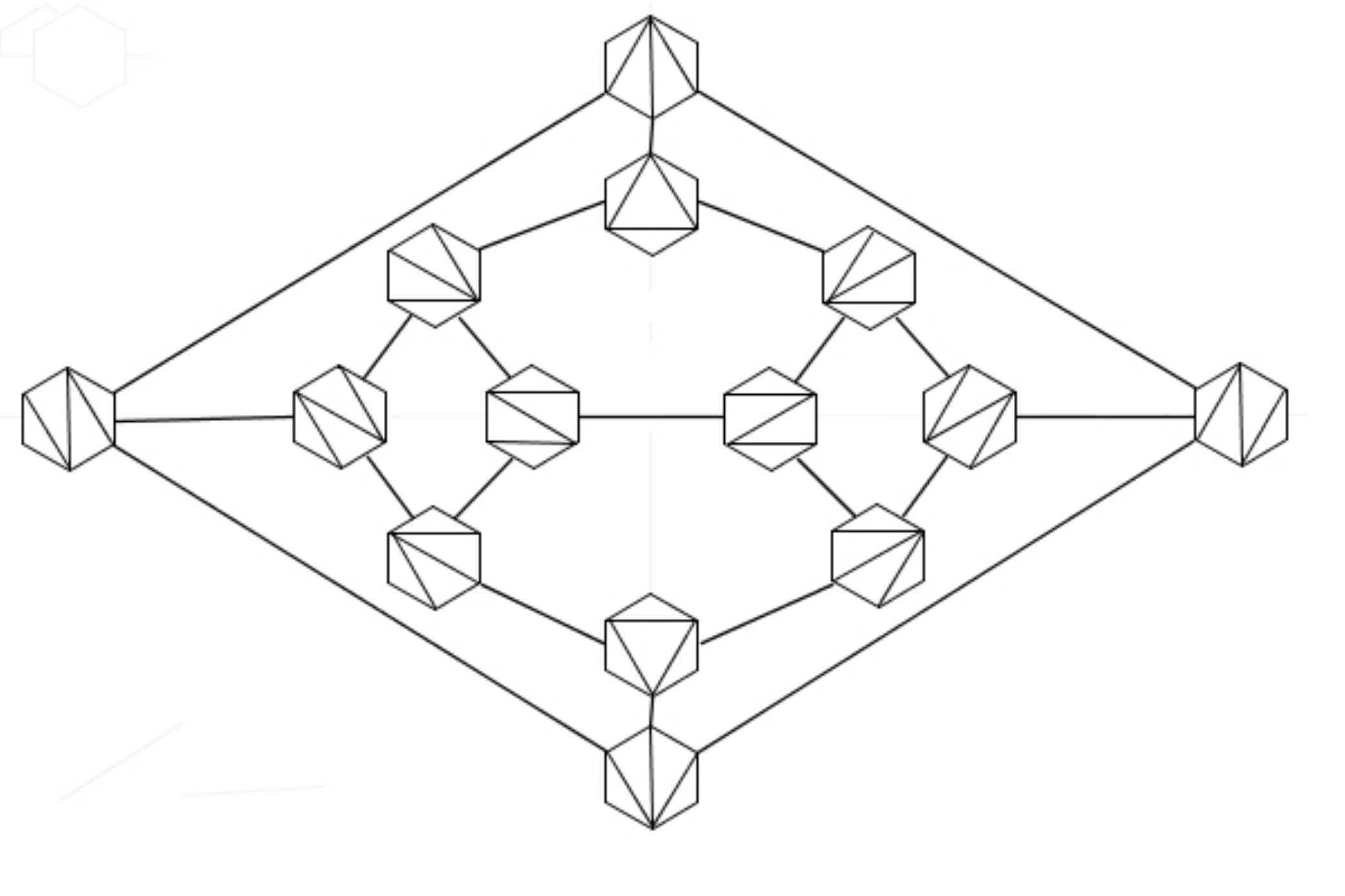}
\end{center}
\caption{The flip graph of a hexagon.}
\end{figure}

\section{Shapes and polyhedral subdivisions}
Given a genotype space $\Sigma^L$,
consider all possible shapes induced by
fitness landscapes 
$
w:\Sigma^L \mapsto \R.
$
We will describe how the shapes are related. 
Most results depend
on triangulations of polytopes.
In particular, we will
discuss the secondary polytope \citep{gkz},
an important construction in discrete 
mathematics. 
The secondary polytope 
is useful for a global understanding of 
shapes.  We will not provide
proofs, but rather
describe results and how
they apply to 
epistasis. For a
thorough  treatment, 
including proofs, 
see \citet{drs} 
and for the biological
perspective \citet{bps}.
Remark \ref{ptheory}
in the end of Section 9 
explains how our applications
relate to the general theory
about polytopes.
Some definitions
below may seem technical,
but the figures and 
intuitive descriptions
from the previous 
sections should help.

Throughout the section,
let ${\bf{A}} \in \R^d$ denote a finite point set.
A {\emph{polytope}} is the convex hull of a point set,
and $\text{conv}({\bf{A}})$ denotes the convex hull of
${\bf{A}}$.

Polytopes include points, line segments, triangles and tetrahedra,
as well as $L$-cubes and polygons. 
We will use some concepts expressed in terms of point sets,
although we have polytopes in mind. 
In particular, a triangulation of
a polytope is a triangulation of the set of its vertices, 
and similarly for the other concepts. 

A $k$-simplex is the convex hull of $k+1$ 
affinely independent points. In particular, 
points, segments, triangles and tetrahedra
are simplices.

A j-{\emph{face}} of
a $k$-simplex is the convex hull of
a subset of $j$ vertices.

We will give a formal
definition of triangulations
and some related concepts.
A {\emph{polyhedral subdivision}} 
of a point set ${\bf{A}}$ is a collection
of polytopes $\mathcal{C}$, such that 
\begin{enumerate}
\item[(i)] If $C \in \mathcal{C}$ then each face of $C$ belongs to 
$\mathcal{C}$ as well (closure property),
\item[(ii)] the union 
$\cup_{C \in \mathcal{C}} C= \text{conv}({\bf{A}})$ (union property),
\item[(iii)] for $C \neq C'$ where $C, C' \in  \mathcal{C}$, 
the intersection $C\cap C'$ does not contain
any interior points of $C$ or $C'$ (intersection property).
\end{enumerate}
A {\emph{triangulation}}
of ${\bf{A}}$
is a polyhedral subdivision
such that all polytopes are simplices.
A {\emph{refinement}} $\mathcal{C'}$ of a polyhedral 
subdivision $\mathcal{C}$ is a polyhedral subdivision 
$\mathcal{C'}$
where for each $C' \in  \mathcal{C'}$,
there exists a $C \in \mathcal{C}$,
such that $C' \subset C$.
A polyhedral subdivision is an {\emph{almost triangulation}}
if it is not a triangulation,
but all its proper refinements are triangulations.
Two triangulations of the
same point set are connected
by a {\emph{flip}}
if they are the only two triangulations
refining an almost triangulation.
All these concepts are illustrated in
Fig. 9 and 13. Specifically, Fig. 13 shows a polyhedral 
subdivision which is also an almost triangulation.
Moreover, the two possible refinements
are the triangulations in Fig. 9.
As mentioned, the triangulations in Fig. 9
differ by a flip, so that the formal
definition agrees with the descriptions
in the previous section.

\begin{figure}
\begin{center} 
\includegraphics[scale=0.5]{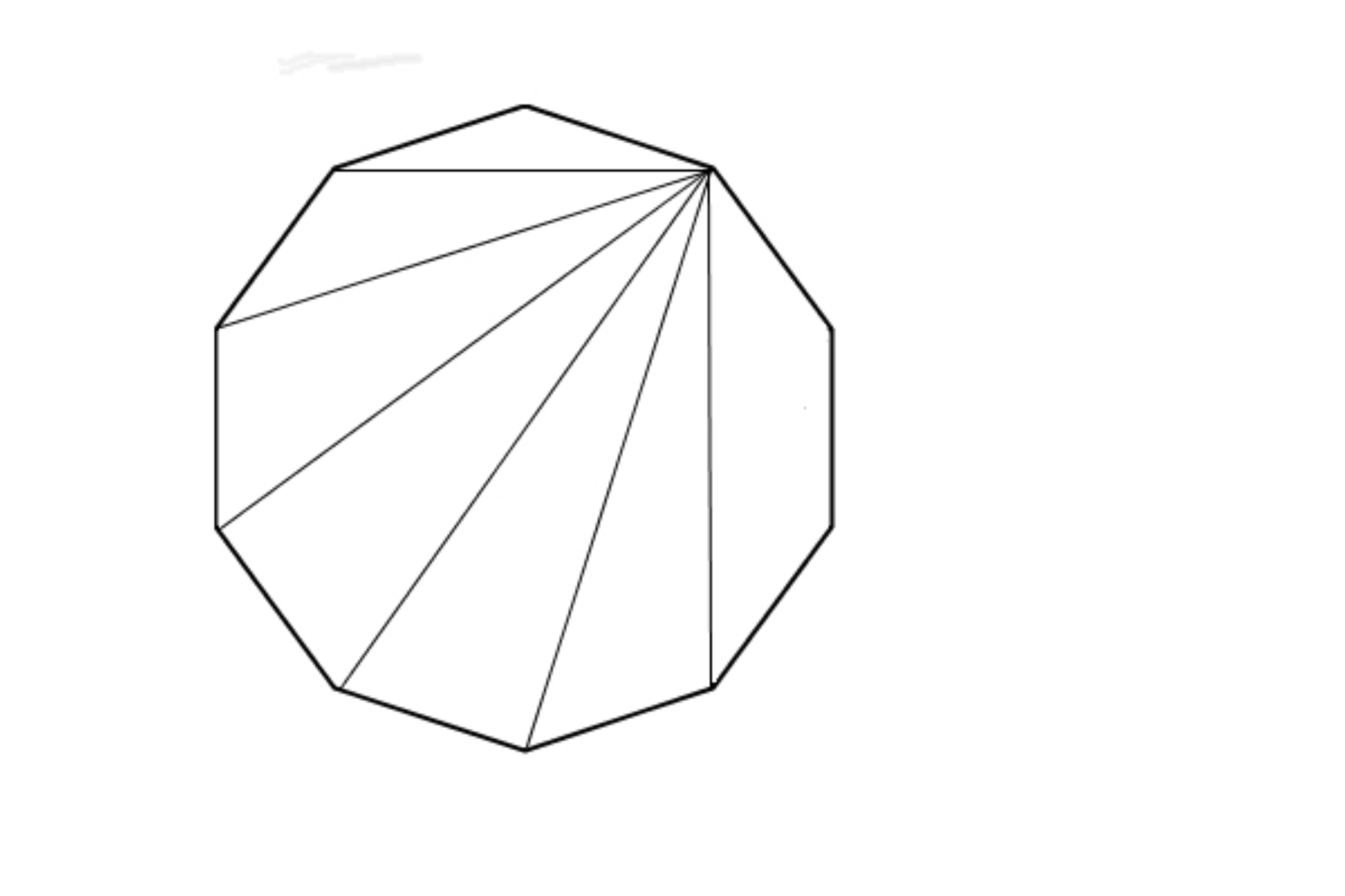}
\end{center}
\caption{The polyhedral subdivision is an
almost triangulation. The two possible refinements
result in the two triangulations from Fig. 9.}
\end{figure}

From the previous section,
a triangulation induced by 
the fitness landscape
is the shape of the landscape.
More generally, we can
describe all shapes
using the concepts defined in
this section.
Consider again
the parametric linear programming
problem where $v$ varies:
\[
\tilde{w} (v) = \max \{ \, p \cdot w : \rho (p)=(v)
{\text{ for all }} v \in [0,1]^L \}.
\]
The domains of linearity of $\tilde{w}$
constitute a polyhedral subdivision
\citep[Chapter 2]{drs} of the genotope.
The {\emph{shape}}
of a fitness landscape is
the polyhedral division
induced by the landscape.
This subdivision is not always a triangulation.
Recall from the two-loci case that
positive epistasis corresponds to
$u>0$ and negative epistasis to $u<0$,
for
\[
u=w_{00}-w_{01}-w_{10}+ w_{11}.
\]
However, one does not obtain a triangulation
if $u=0$. As remarked,
the fitness landscape is generic
if it induces a triangulation,
and the corresponding shapes are
the generic shapes.

In order to further describe the
relations between shapes, we will 
consider minimal dependence sets
of points, as in the following example.
\begin{example}
Consider the vertices
of the genotope $[0,1]^2$.
The relation
\[
1\cdot (0,0)- 1 \cdot(0,1)
-1 \cdot (1,0)+ 1\cdot (1,1)=0,
\]
is an affine dependence relation,
since the sum of the coefficients $1-1-1+1$
is zero.
This set of four points is a minimal
dependence set, in the sense
that every proper subset of the four points
is independent.
The form
\[
w_{00}-w_{01}-w_{10}+ w_{11},
\]
corresponds to the dependence
relation. Notice that the form is
unique up to scaling, i.e., multiplication
by a constant. 
\end{example}

We define a {\emph{circuit}} as
a minimal affine dependence set.
The corresponding forms are called
circuits as well, and they 
are unique up to scaling.
Flips and circuits are closely related.
The circuit 
$u$  
corresponds to the flip between the two
triangulations of 
the genotope in the two-loci case. 
More precisely,
the triangulation corresponding
to positive epistasis is described
by $u>0$, and the flip corresponds
to replacing $u>0$ by $u<0$.
In general, a flip corresponds
to changing sign of a circuit,
and some examples for $L=3$ are given
in the next section.
The next concept will be used for 
defining the secondary polytope,
and for describing flips and circuits
in more detail.

For a triangulation of a polytope, we define
the {\emph{GKZ vector}} as follows:
Each component of the GKZ vector
corresponds to a vertex of the
polytope. The component
is the sum of the normalized
volumes of all simplices containing 
the vertex. 

In this context, it is sufficient
to use that the normalized volume of 
the $L$-cube $[0,1]^L$ is $n!$,
so that the genotope $[0,1]^2$ 
in the two loci case has area 2,
and the genotope
$[0,1]^3$ in the 3-loci case
has volume 6. In general, for an n-dimensional lattice
polytope, the normalized volume is defined as
the Euclidean volume of the polytope
multiplied by $n!$

\begin{example}
The GKZ-vector
for the triangulation of $[0,1]^2$
in the two-loci
case associated with positive epistasis
is $(2,1,1,2)$, using the order as 
given in the beginning of Section 6.
Indeed, both triangles have area 1 (using
normalized volumes).
The vertices $00$ and $11$
belong to two different triangles each,
whereas $10$ and $01$ are "sliced off",
so that each of them
belong to one triangle only.
Similarly, the
GKZ-vector
for the triangulation
associated with negative epistasis
is $(1,2,2,1)$.
\end{example}

The purpose with the next example is
to relate circuits and GKZ vectors.
\begin{example}
From the previous
example, the GKZ-vector 
for the triangulation
associated with positive epistasis
is $(2,1,1,2)$, whereas the
the GKZ vector for the triangulation
associated to negative epistasis
is $(1,2,2,1)$.
We relate to the circuit
$u=w_{00}-w_{01}-w_{10}+ w_{11}$ 
the vector $(1,-1,-1,1)$.
The flip between the two
triangulations
corresponds to $u$,
and
\[
(2,1,1,2)-(1,2,2,1)=(1,-1,-1,1)
\]
so that the GKZ vectors
differ by the vector 
corresponding to the circuit associated
with the flip.
\end{example}
See Remark \ref{ptheory}
for some comments about
the relation between 
GKZ vectors and flips,
including references.
In the next section, we will
consider the relations
between flips
and GKZ vectors for $L=3$.

For a given polytope
the {\emph{secondary polytope}} is 
defined as the
the convex hull of the GKZ vectors.
The geometric classification of
fitness landscapes depends on the 
secondary polytope.
For a genotope,
the vertices of the secondary polytope
correspond to generic shapes,
and its edges to flips between the
generic shapes.
The higher dimensional
faces of the secondary
polytope correspond to
non-generic shapes.
Consequently, the secondary polytope
represent all the shapes
and their relations.

\begin{example}
The secondary polytope for the two-loci case is a line
segment, where the vertices
corresponds to the two triangulations, 
and the line segment to the flat shape.
\end{example}

\newpage
\section{The 74 generic shapes of the cube}
The relations between shapes, 
circuits and GKZ vector for the 3-cube,
is analogous to the two-loci
case, as indicated
in the previous section.
Recall that the square has 2 generic shapes, 
corresponding to $u>0$ and $u<0$,
for
\[
u=w_{00}-w_{01}-w_{10}+ w_{11}.
\]
The cube has 74 generic 
shapes, where a shape is determined by 
the following 20 circuits:

\begin{align*} 
a&:= w_{000}-w_{010}-w_{100}+w_{110}\\ 
b&:=w_{001}-w_{011}-w_{101}+w_{111}\\
c&:=w_{000}-w_{001}-w_{100}+w_{101}\\
d&:=w_{010}-w_{011}-w_{110}+w_{111}\\
e&:=w_{000}-w_{001}-w_{010}+w_{011}\\
f&:=w_{100}-w_{101}-w_{110}+w_{111}\\
g&:=w_{000}-w_{011}-w_{100}+w_{111}\\
h&:=w_{001}-w_{010}-w_{101}+w_{110}\\
i&:=w_{000}-w_{010}-w_{101}+w_{111}\\
j&:=w_{001}-w_{011}-w_{100}+w_{110}\\
k&:=w_{000}-w_{001}-w_{110}+w_{111}\\
l&:=w_{010}-w_{011}-w_{100}+w_{101}\\
m&:=w_{001}+w_{010}+w_{100}-w_{111}-2w_{000}\\
n&:=w_{011}+w_{101}+w_{110}-w_{000}-2w_{111}\\
o&:=w_{010}+w_{100}+w_{111}-w_{001}-2w_{110}\\
p&:=w_{000}+w_{011}+w_{101}-w_{110}-2w_{001}\\
q&:=w_{001}+w_{100}+ w_{111}-w_{010}-2w_{101}\\
r&:=w_{000}+w_{011}+ w_{110}-w_{101}-2w_{010}\\
s&:=w_{000}+w_{101}+ w_{110}-w_{011}-2w_{100}\\
t&:=w_{001}+w_{010}+w_{111}-w_{100}-2w_{011}
\end{align*} 
We will use the letters $a-t$ in the list,
as well as $u$, throughout the section.

In order to emphasize the
connection to gene interactions,
especially the algebraic
aspects of epistasis,
we will consider the
{\emph{interaction space}}.
For any $L$,
let $\mathcal{L}$ be the subspace of $\R^{\Sigma^L}$
consisting of additive fitness landscapes.
The interaction space
is the vector space dual
to the quotient of
$\R^{\Sigma^L}$ modulo $\mathcal{L}$, 
or
\[
{
\left(
\R^{\Sigma^L}/\mathcal{L}
\right)
}^\ast .
\]
The interaction space
is spanned by the set of 
all circuits, where the
circuits are unique up to 
scaling. In the two-loci case,
the interaction space
is spanned by $u$.
For $L=3$, the interaction space
is spanned by the
20 circuits $a-t$.

The {\emph{circuit sign pattern}} of
a fitness landscape 
consists of the sign (positive, negative 
or zero) of each circuit. 
In the two-loci case there is only one 
circuit and the sign 
pattern is either $u>0$, $u<0$ or $u=0$.
A central result for the geometric classification
is that the circuit sign pattern determines the
shape of the fitness landscape, but in
general the converse does not hold \citep{bps}. 
In particular,
the signs of the 20 circuits $a-t$
determine the shape of the fitness landscape
for $L=3$.
In total, there are 74 generic shapes.
The fact that there are 20 circuits and
only 74 generic shapes reflects
dependence relations.

Table 1 lists the shapes,
where the vertices of the cube
are ordered as  follows
\[
000, 001, 010, 011,  100, 101, 110, 111.
\]

\begin{table}
\caption{Shape numbers, GKZ vectors, inequalities and adjacent shapes}
\centering
\begin{tabular}{ l | l | l || l |l | l } 
$\#$ & GKZ  & inequalities &  $\#$ & GKZ & circuits \\
\hline
1 &15515115 & tqom3,4,5,6 
&38 &31355313 &$\overline{\text{l}} \overline{\text{g}}$cd39,44,51,59 \\
2 &51151551 &srpn7,8,9,10
&39 &31533513 &l$\overline{\text{i}}$ef38,44,53,60 \\
3 &14436114 & $\overline{\text{t}}$bd$\overline{\text{e}}$1,11,13,17 
&40 &33155133 & $\overline{\text{j}} \overline{\text{g}}$ab42,45,54,61 \\
4 &14614314 &  $\overline{\text{q}}$bf$\overline{\text{c}}$1,12,14,18
&41 &33511533 &$\overline{\text{h}} \overline{\text{i}}$ab43,46,55,62\\
5 &16414134 &$\overline{\text{o}}$df$\overline{\text{a}}$1,15,16,19 
&42 &35133153 &j$\overline{\text{k}}$ef40,45,57,63\\
6 &34414116 
&$\overline{\text{m}} 
\overline{\text{e}}
\overline{\text{c}}\overline{\text{a}}$1,28,29,31
&43 &35311353 &h$\overline{\text{k}}$cd41,46,58,64\\
7 &41163441 & $\overline{\text{s}}$ac$\overline{\text{f}}$2,20,22,26
&44 &51333315 & gi$\overline{\text{b}} \overline{\text{a}}$38,39,65,68\\
8 &41341641 &$\overline{\text{r}}$ae$\overline{\text{d}}$2,21,23,27
&45 &53133135 &gk$\overline{\text{d}} \overline{\text{c}}$40,42,66,69\\
9 &43141461 &$\overline{\text{p}}$ce$\overline{\text{b}}$2,24,25,30  
&46 &53311335 &ik$\overline{\text{f}}\overline{\text{e}}$41,43,67,70\\
10 &61141443 
&$\overline{\text{n}}\overline{\text{f}}
\overline{\text{d}}\overline{\text{b}}$2,32,33,34 
&47 &13356222 
&$\overline{\text{d}}\overline{\text{b}}$f$\overline{\text{e}}$11,13,35,71\\
11 &13446213 
&$\overline{\text{b}}\overline{\text{l}}$d$\overline{\text{e}}$3,12,47,51
&48 &13623522 
&$\overline{\text{f}}\overline{\text{b}}$d$\overline{\text{c}}$12,14,36,72\\
12 &13624413 &$\overline{\text{b}}$lf$\overline{\text{c}}$4,11,48,53
&49 &16323252 & $\overline{\text{f}}\overline{\text{d}}$b$\overline{\text{a}}$15,16,37,73\\
13 &14346123 
&$\overline{\text{d}}\overline{\text{j}}$b$\overline{\text{e}}$3,15,47,54 
&50 &22265331 
&$\overline{\text{c}}\overline{\text{a}}$e$\overline{\text{f}}$20,22,35,71\\
14 &14613423 
&$\overline{\text{f}}\overline{\text{h}}$b$\overline{\text{c}}$4,16,48,55
&51 &22356213 &e$\overline{\text{b}}\overline{\text{c}}$d11,17,38,71\\
15 &16324143 &$\overline{\text{d}}$jf$\overline{\text{a}}$5,13,49,57
&52 &22532631 
&$\overline{\text{e}}\overline{\text{a}}$c$\overline{\text{d}}$21,23,36,72\\
16 &16413243 &$\overline{\text{f}}$hd$\overline{\text{a}}$5,14,49,58
&53 &22623513 &c$\overline{\text{b}}\overline{\text{e}}$f12,18,39,72\\
17 &23346114 &e$\overline{\text{g}}$bd3,28,51,54
&54 &23256123 &e$\overline{\text{d}}\overline{\text{a}}$b13,17,40,71\\
18 &23613414 &c$\overline{\text{i}}$bf4,29,53,55
&55 &23612523 &c$\overline{\text{f}}\overline{\text{a}}$b14,18,41,72\\
19 &26313144 &a$\overline{\text{k}}$df5,31,57,58
&56 &25232361 
&$\overline{\text{e}}\overline{\text{c}}$a$\overline{\text{b}}$24,25,37,73\\
20 &31264431 
&$\overline{\text{a}}\overline{\text{l}}$c$\overline{\text{f}}$7,21,50,59
&57 &26223153 &a$\overline{\text{d}}\overline{\text{e}}$f15,19,43,73\\
21 &31442631 &$\overline{\text{a}}$le$\overline{\text{d}}$8,20,52,60
&58 &26312253 &a$\overline{\text{f}}\overline{\text{c}}$d16,19,43,73\\
22 &32164341 
&$\overline{\text{c}}\overline{\text{j}}$a$\overline{\text{f}}$7,24,50,61
&59 &31265322 &f$\overline{\text{a}}\overline{\text{d}}$c20,26,38,71\\
23 &32431641 
&$\overline{\text{e}}\overline{\text{h}}$a$\overline{\text{d}}$8,25,52,62
&60 &31532622 &d$\overline{\text{a}} \overline{\text{f}}$e21,27,39,72\\
24 &34142361 &$\overline{\text{c}}$je$\overline{\text{b}}$9,22,56,63
&61 &32165232 &f$\overline{\text{c}}\overline{\text{b}}$a22,26,40,71\\
25 &34231461 &$\overline{\text{e}}$hc$\overline{\text{b}}$9,23,56,64 
&62 &32521632 &d$\overline{\text{e}}\overline{\text{b}}$a23,27,41,72\\
26 &41164332 &f$\overline{\text{g}}$ac7,32,59,61
&63 &35132262 &b$\overline{\text{c}}\overline{\text{f}}$e24,30,42,73\\
27 &41431632 &d$\overline{\text{i}}$ae8,33,60,62
&64 &35221362 &b$\overline{\text{e}}\overline{\text{d}}$c25,30,32,73\\
28 &43324116 &eg$\overline{\text{c}}\overline{\text{a}}$ 6,17,65,66
&65 &52323216 &ceb$\overline{\text{a}}$28,29,44,74\\
29 &43413216 &ci$\overline{\text{e}}\overline{\text{a}}$6,18,65,67
&66 &53223126 &aed$\overline{\text{c}}$28,31,45,74\\
30 &44131362 &b$\overline{\text{k}}$ce9,34,63,64
&67 &53312226 &acf$\overline{\text{e}}$29,31,46,74\\
31 &44313126 &ak$\overline{\text{e}}
\overline{\text{c}}$6,19,66,67
&68 &61232325 &dfa$\overline{\text{b}}$32,33,44,74\\
32 &61142334 &fg$\overline{\text{d}} \overline{\text{b}}$10,26,68,69
&69 &62132235 &bfc$\overline{\text{d}}$32,34,45,74\\
33 &61231434 &di$\overline{\text{f}} \overline{\text{b}}$10,27,68,70
&70 &62221335 &bde$\overline{\text{f}}$33,34,46,74\\
34 &62131344 &bk$\overline{\text{f}}\overline{\text{d}}$10,30,69,70
&71 &22266222 
&ef$\overline{\text{d}}\overline{\text{b}}
\overline{\text{c}}\overline{\text{a}}$47,50,51,54,59,61\\
35 &13355331 
&$\overline{\text{l}}\overline{\text{j}}
\overline{\text{f}}\overline{\text{e}}$36,37,47,50 
&72 &22622622 
&cd$\overline{\text{f}}\overline{\text{b}}
\overline{\text{e}}\overline{\text{a}}$48,52,53,55,60,62\\
36 &13533531 &l$\overline{\text{h}} 
\overline{\text{d}}\overline{\text{c}}$35,37,48,52
&73 &26222262 &ab$\overline{\text{f}}
\overline{\text{d}}\overline{\text{e}}\overline{\text{c}}$49,56,57,58,63,64\\
37 &15333351 &jh$\overline{\text{b}}\overline{\text{a}}$35,36,49,56
&74 &62222226 &acebdf65,66,67,68,69,70
\end{tabular}
\end{table}

For each shape, the table gives
the GKZ vector, the defining inequalities,
and the adjacent shapes.
In particular, for Shape 74 
the notation
\[
\text{acebdf}65,66,67,68,69,70,
\]
in the second column
means that Shape
74 is defined by 
\[
a, c, e, b, d ,f > 0.
\]  
and that the adjacent 
shapes
are 65,66,67,68,69,70.
Similarly,
for Shape 65
the notation
\[
\text{ceb} \overline{\text{a}}28,29,44,74,
\]
means that the shape is defined by
\[
c>0,e>0, b>0, a<0,
\]
where $\overline{\text{a}}$
indicates that $a<0$,
and the adjacent
shapes are 28, 29, 44, 74.

Each inequality of Shape 74
can be described in terms
of epistasis (in the usual sense),
since each inequality keeps one
locus fixed. In contrast, the
inequalities of Shape 1 considers
three-way interactions. 
The fact that  $m>0$,
where
\[
m=w_{001}+w_{010}+w_{100}-w_{111}-2w_{000}
\]
shows that the genotype 111
has lower fitness  as compared to
a linear expectation
from the values  
\[
w_{001}, w_{010}, w_{100},  w_{000}.
\]
This observation shows already
that the geometric theory is more
fine-scaled as compared to conventional
approaches.

The 74 shapes fall into six categories, called
the {\emph{interaction types}}. 
Specifically, the types
consist of Shape 1-2, 3-10, 11-34, 35-46, 47-70 and 71-74.
For pictures of the six interaction types, see
\citet[Chapter 1]{drs}.
Shapes of the same type differ only in the labeling of the vertices. 
In particular, the shapes of the 
same interaction type in the table have GKZ vectors that 
differ only by a permutation of the components.

As in the two-loci case,
the circuits correspond
to flips. The letters representing
circuits are ordered
according to the shapes
resulting from the
corresponding flips.
Consider again Shape 74.
In addition to the
information described,
the notation
\[
acebdf65,66,67,68,69,70.
\]
indicates how flips and
shapes relate in a precise
way. The flip corresponding to $a$
results in Shape 65 (the first letter 
is paired with the first number), 
the flip corresponding to
$c$ results in Shape 66, and so forth.

For an explicit description, consider
Shape 74 and the flip corresponding
to $a$. Since Shape 74 is defined by
\[
a>0, c>0, e>0, b>0, d>0, f>0,
\]
the result of the flip is the
shape defined by
\[
a<0, c>0, e>0, b>0, d>0, f>0,
\] 
which reduces to
\[
a<0, c>0, e>0, b>0,  
\]
since the four
inequalities imply 
that $d>0$ and  $f>0$.
Shape 65 is 
described by
exactly these four inequalities
in the table.

The table lists GKZ vectors
as well. Flips and
GKZ-vectors  
are related,
as in the two-loci case.
For instance, the  GKZ vector 
is 62222226 for Shape 74,  
and 52323216 for Shape 65.
The circuit $a$ corresponds to
the vector $(1,0,-1,0,-1,0,1,0)$,
and
\[
(6,2,2,2,2,2,2,6)-(5,2,3,2,3,2,1,6)=
(1,0,-1,0,-1,0,1,0).
\]
(see Remark \ref{ptheory}).
For a systematic interpretation of
the 20 circuits $a-t$ listed, one may
consider the Fourier transform for
the group ${(\Z_2)}^n$ \citep{bps}.
Geometric interpretations of the
circuits are given in the same paper.

\begin{remark}\label{ptheory}
We have indicated results
about polytopes and triangulations
of relevance in evolutionary biology (see also 
the next section about shapes and empirical data).
We refer to \citet{drs}
for general background
about triangulations of
polytopes, where the Chapters 4 and 5
are especially important.
"Additive fitness landscapes", as defined here,
translates to "linear evaluations"
in the general theory,
and "interaction spaces" to 
"linear dependences".
The fact that the interaction space
is spanned by the circuits is 
an aspect of Gale duality 
\citep[Chapter 4]{drs}.
The relation between GKZ
vectors and flips were illustrated
above for the Shapes 65 and 74,
and in Example 8 in the previous section. 
In terms of the general theory,
the interaction space equals
the linear space parallel 
to the secondary polytope
\citep[Chapter 5]{drs},
and a detailed description
of the relation between 
GKZ vectors and
flips 
is given in the 
same chapter.
\end{remark}

\section{Shapes and empirical data}
The described
relations between circuits, flips,
GKZ-vectors and the
secondary polytopes
hold under very
general assumptions.
We restricted our discussion
to biallelic $L$-loci
populations in order to
keep the presentation 
simple.
For the geometric theory of gene interactions, 
the genotope is defined for any set of genotypes found in a
population and the shape is defined accordingly
\citep{bps}. In fact, the authors stress that the genotope is never an 
$L$-cube for binary data and many loci ($\geq 20$),
which is important for complexity reasons.
Empirical examples of general type
\citep{bps, bpse} can be analyzed 
similarly to the restricted case we considered
here.
For a shape analysis of empirical data, 
one needs several fitness measurements
of each genotype due to statistical 
variation. One may not find a unique shape,
but rather a set of similar shapes which 
are compatible with the data. 

A shape analysis
of HIV fitness data is given in \citep{bps}. 
The biallelic three-loci system considered 
there is associated with HIV drug resistance. 
From bootstrapping, the three dominant shapes
are $2$, $7$ and $10$. Notice that these shapes
Ëare adjacent, and have similar GKZ vectors.
Moreover, the five most dominant shapes 
$2$, $7$, $10$, $26$ and $32$ appear 
as a face of the secondary polytope
of the cube, and have similar GKZ vectors
as well.

Software for analyzing shapes is available,
for example Polymake

(http://www.polymake.org/doku.php).

\section{Shapes and fitness graphs}
A fitness graph is determined by
fitness ranks only.
The information from
shapes is incomparably more
fine-scaled. It is of interest
to compare the two perspectives.

For the two-loci case, assume that the
$11$ genotype has maximal fitness.
Then positive epistasis is compatible
with three
fitness graphs (no arrows down, exactly one arrow down, or two arrows
down). On the other hand, consider the fitness graphs with all arrows
up. Such a graph is compatible with positive, negative or
no epistasis. This example shows that fitness graphs provide
information that cannot be obtained from the geometric classification,
and vice versa, and the same observation holds for any $L$.
Since $L$=2 is rather special, we will demonstrate that fitness graphs
and shapes provide complementary information also
for $L=3$, where the examples are from \citet{cgb}.
For more comparisons of fitness graphs and shapes, also in the context of
empirical data, see the same paper.

The following all arrows up landscapes is of Shape 74,
\[
w_{000}=1, w_{100}=w_{010}=w_{001}=2,
w_{110}=w_{101}=w_{011}=4, w_{111}=7.
\]
The landscape
\[
w_{000}=2, w_{100}=w_{010}=w_{001}=1,
w_{110}=w_{101}=w_{011}=4, w_{111}=8,
\]
is of shape 74 as
well. The corresponding fitness
graph has exactly 3 arrows down,
and both $000$ and $111$
are at peaks.
It is easily seen
that there exist
fitness landscapes
of shape 74 with
other fitness graphs,
in addition to the two
mentioned.

For each interaction type for $L=3$,
Table 2 gives the shape number
and an example of an
all arrows up landscape of this
shape.

\begin{table}
\begin{center}
\caption{Interaction types, shape numbers and fitness landscapes}
\begin{tabular}{clccccccc}
& $w_{000}$ & $w_{100}$  & $w_{010}$  & $w_{001}$  & $w_{110}$  &
$w_{101}$
&$w_{011}$  & $w_{111}$ \\
\hline
Type 1, no 2: & 1 & 2 & 2 & 2 & 4 & 4 & 4 & 5  \\
Type 2, no 10: &1 & 2 & 2 & 2 & 6 & 6 & 6 & 9 \\
Type 3, no 34: &1 & 2 & 2 & 2 & 10 & 6 & 5 & 12\\
Type 4, no 46:  &1 & 2 & 5 & 5 & 8 & 8 & 8 & 13 \\
Type 5, no 70:  &1 & 2 & 5 & 5 & 9 & 9 & 10 & 15 \\
Type 6, no 74: &1 & 2 & 2 & 2 & 4 & 4 & 4 & 7 \\
\end{tabular}
\end{center}
\end{table}

As we have seen, fitness graphs and
shapes provide complementary information.
There is usually an overlap in the
information as well. For instance,
if all arrows point away from
a particular genotype,
and if the genotype is "sliced
off" for the shape, then one
has two indications that
the genotype has relatively low fitness.

Fitness graphs
provide information
about the adaptive potential
if one restricts to
(single) mutations.
The graphs reveal
coarse properties,
such as sign epistasis,
mutational trajectories,
and the number of peaks.
It is clear that a
complete analysis
of recombination
requires more fine-scaled
information as compared to
what fitness graphs provide.
The geometric theory, on
the other hand,
reveals all gene interactions.
Finding the shape of a
landscape is equivalent to finding
all the fittest populations,
which explains why shapes
are relevant for recombination.

From a more philosophical perspective, the interest in fitness graphs and 
shapes
depends on the belief that average effects of mutations are insufficient for
analyzing evolutionary dynamics.

\section{Discussion}
We have considered
fitness graph
and the geometric theory
of gene interactions.
Fitness graphs and shapes
provide complementary information,
and there tend to be some overlap
in the observations.
Fitness graphs
are useful for
analyzing
peaks,
and other coarse 
properties of
fitness landscapes.
The graphs have
been used in empirical work,
and for relating global
and local properties of fitness
landscapes.  

The geometric theory extends 
the usual concept epistasis to 
any number of loci, 
where shapes, as defined in the
geometric theory, correspond
to positive and negative
epistasis for two mutations.
The geometric classification
is meaningful because it comes with
a structure. A particular shape
can be put in a context, and compared
to other shapes. In summary, for biallelic 
populations where all $2^L$ genotypes are 
represented, the genotope is an $L$-cube. 
The shape of a fitness landscape
is a polyhedral subdivision of
the genotope induced by the landscape.
The generic shapes are the 
triangulations of the genotope. 
The relation between all the generic shapes 
can be described in terms of flips, or minimal
changes between shapes. 
The flip graph provides an overview 
of the generic shapes and how they can 
be transformed into each other by flips. 
The secondary polytope encodes all 
shapes and their relations, where the generic 
shapes correspond to vertices,
and the non-generic shapes to
the higher dimensional faces.
For an algebraic perspective,
the interaction space is spanned 
by a set of linear forms, or circuits.
The shapes are determined by the sign 
pattern of the circuits, and changing 
sign of a circuit corresponds to a flip. 

The geometric theory has 
provided new insights  
about gene interactions in
empirical studies.
The theory may be considered a
fundamentally new approach
to recombination. 
There is clearly a potential for new 
applications of shapes to evolutionary 
biology as well as various empirical 
problems, even if the theory is
complete.

The approaches discussed here
are similar in one respect. They make
no assumptions, or minimal assumptions,
about the underlying fitness
landscapes. The accuracy of an analysis
of empirical data using fitness graphs
or the geometric theory 
does not depend on any {\emph{a priori}}
assumptions about the fitness 
landscape. Fitness graphs and shapes
are well suited for empirical 
studies for that reason.


\begin{thebibliography}{99}

\bibitem[Aita et al., 2001]{aih}
Aita, T., Iwakura, M. and Husimi, Y. (2001).
A cross-section of the fitness landscape of dihydrofolate reductase.
{\emph{Protein Eng.}} Sep; 14(9):633--8.

\bibitem[Beerenwinkel et al., 2007 a]{bes}  
Beerenwinkel, N., Eriksson, N. and Sturmfels, B. (2007).
Conjunctive Bayesian networks.
{\emph{ Bernoulli}}; 13:893--909.

\bibitem[Beerenwinkel et al., 2007 b]{bps}
Beerenwinkel, N., Pachter, L. and Sturmfels, B. (2007).
Epistasis and shapes of fitness landscapes.
{\emph{Statistica Sinica}} 17:1317--1342.


\bibitem[Beerenwinkel et al., 2007 c]{bpse}
Beerenwinkel, N., Pachter, L., Sturmfels, B., Elena, S. F. and
Lenski, R. E. (2007). Analysis of epistatic interactions and fitness
landscapes using a new geometric approach.
{\emph{BMC Evolutionary Biology}} 7:60.


\bibitem[Carnerio and Hartl, 2010]{ch}
Carnerio, M. and Hartl, D. L. (2010). Colloquium papers: Adaptive
landscapes and
protein evolution. Proc. Natl. Acad. Sci USA 107 suppl 1: 1747-1751.


\bibitem[Crona et al., 2013 a]{cgb}
Crona, K., Greene, D. and Barlow, M. (2013).
The peaks and geometry of fitness landscapes.
{\emph{J. Theor. Biol.}} 317: 1--13.

\bibitem[Crona et al., 2013 b]{cps}
Crona, K., Patterson, D.
Stack K.,  Greene D.,
Goulart, C. P., Mentar, M.,
Jacobs, S. J., Kallmann, M.
and Barlow, M. (2013).
Antibiotic resistance landscapes:
a quantification of
theory-data incompatibility
for fitness landscapes.
http://arxiv.org/abs/1303.3842

\bibitem[Crow, 2006]{crow}
Crow, J. F. (2006).
H. J. Muller and the "competition hoax"
{\emph{Genetics.}} vol. 173 no. 2 511--514

\bibitem[De Loera et al., 2010]{drs}
De Loera, J. A., Rambau, J. and Santos, F. (2010).
Triangulations: Applications,
Structures and Algorithms. Number 25 in Algorithms and Computation in
Mathematics. Springer-Verlag, Heidelberg.

\bibitem[De Visser et al., 2009]{dpk}
De Visser, J. A. G. M., Park S.C. and
Krug J.( 2009).
Exploring the effect of sex on empirical fitness landscapes.,
{\emph{The American Naturalist}}.


\bibitem[Desper et al., 1999]{djk}
Desper, R., Jiang, F., Kallioniemi, O.P., Moch, H., Papadimitriou, C.H.
and Sch\"affer, A.A. (1999).
Inferring tree models for oncogenesis from comparative genome
hybridization data.
{\emph{Comput. Biol}} 6 37--51.

\bibitem[Flyvberg and Lautrup, 1992]{fl}
Flyvbjerg, H. and Lautrup, B. (1992). Evolution in a rugged fitness
landscape.
Phys Rev A 46:6714-6723.

\bibitem[Franke et al., 2011]{fkd}
Franke, J.,  Kl{\"o}zer, A., de Visser, J.A.G.M.
and Krug., J. (2011). Evolutionary Accessibility
of Mutational Pathways.
{\emph{PLoS Comput Biol}} 7(8):
e1002134. doi:10.1371/journal.pcbi.1002134.


\bibitem[Gelfand et al., 1994]{gkz}
Gelfand, I. M., Kapranov, M. and Zelevinsky, A. (1994).
Discriminants, resultants and multidimensional
determinants, Birkh{\"a}user.

\bibitem[Gilliespie, 1983]{g1983}
Gillespie, J. H. (1983). A simple stochastic gene substitution model.
{\emph{Theor. Pop. Biol.}} 23 : 202--215.

\bibitem[Gilliespie, 1984]{g1984}
Gillespie, J. H. (1984). The molecular clock may be an episodic
clock. {\emph{Proc. Natl. Acad. Sci. USA 81}} : 8009--8013.


\bibitem[Goulart et al., 2013]{gmc}
Goulart, C. P., Mentar, M., Crona, K., Jacobs, S. J., Kallmann, M.,
Hall, B. G., Greene D., Barlow M. (2013). Designing antibiotic
cycling strategies by determining and understanding local
adaptive landscapes.
{\emph{PLoS ONE}} 8(2): e56040.
doi:10.1371/journal.pone.0056040.


\bibitem[Haldane, 1931]{h}
Haldane, J. B. S. (1931). Proc. Cambridge Philos. Soc. 27: 37-142


\bibitem[Kauffman and Levin, 1987]{kl}
Kauffman, S. A. and Levin, S. (1987).
Towards a general theory of adaptive
walks on rugged landscapes.
{\emph{J. Theor. Biol}}
128:11--45.

\bibitem[Kauffman and Weinberger, 1989]{kw}
Kauffman, S. A. and Weinberger, E.D. (1989).
The NK model of rugged fitness landscape 
and its application to maturation of the immune response.
{\emph{J. Theor. Biol.}};141:211--245.




\bibitem[Kingman, 1978]{k}
Kingman, J. F. C. (1978). A simple model for the balance between
selection and mutation.
{\emph{J. Appl. Prob.}}
15:1--12.

%




\bibitem[Kryazhimskiy et al., 2011]{kdp}
Kryazhimskiy, S., Draghi, J. A. and Plotkin, J. B. (2011).
In evolution, the sum is less than its part.
{\emph{Science 332}},
1160--1161

\bibitem[Lenski et al., 2003]{lop}
Lenski, R. E., Ofria, C., Pennock, R. T. and Adami, C. (2003). 
The evolutionary origin of complex features.
{\emph{Nature}} 423,
139--144.


\bibitem[Macken and Perelson, 1995]{mp}
Macken, C. A. and Perelson, A. S. (1995).
Protein evolution on partially correlated landscapes.
{\emph{Proc. Natl. Acad. Sci. USA.}} 92:9657--9661.


\bibitem[Mani et. al, 2008]{moh}
Mani, R., St. Onge. R.P., Hartman, J.L., Giaever. G. and Roth. F.P.
(2008).
Defining Genetic Interaction.
{\emph{Proc Natl Acad Sci U S A.}} 105: 3461--3466.

\bibitem[Martin et. al, 2006]{mol}
Martin, G., Otto,. S.P. and Lenormand, T. (2006).
Selection for recombination in structured populations. {\emph{Genetics}}
172:593--609.

\bibitem[Maynard Smith, 1970]{ms}
Maynard Smith, J. (1970). Natural selection and the concept of protein
space. {\emph{Nature}} 225:563--64.



\bibitem[Orr, 2006]{o2006}
Orr, H. A. (2006). The population genetics of adaptation on
correlated fitness landscapes: the block model.
{\emph{Evolution}}; 60:1113--1124.

\bibitem[Otto and Lenormand, 2002]{ol}
Otto, S. P. and Lenormand, T. (2002). Resolving the
paradox of sex and
recombination.
{\emph{Nature Reviews Genetics}}; 3:252-261.


\bibitem[Park and Krug, 2008]{pk}
Park, S. C. and Krug J. (2008). Evolution in random fitness landscapes:
The infinite sites model.
J Stat Mech P04014.

\bibitem[Poelwijk et al., 2007]{pkw}
Poelwijk, F.J., Kiviet, D. J., Weinreich, D. M. and Tans, S.J. (2007).
Empirical fitness landscapes reveal accessible evolutionary paths.
{\emph{Nature}} 445:383--386.

\bibitem[Poelwijk et al., 2011]{psk}
Poelwijk, F. J., Sorin, T.-N., Kiviet, D. J. and Tans, S. J. (2011).
Reciprocal sign epistasis is a necessary condition for
multi-peaked fitness landscapes.
{\emph{J. Theor. Biol.}} Mar 7; 272(1):141--4. 


\bibitem[Rokyta et al., 2006]{rbj}
Rokyta, D. R., Beisel, C. J. and Joyce P. (2006) Properties of adaptive
walks on uncorrelated
landscapes under strong selection and weak mutation. 
{\emph{J Theor. Biol.}}
2006;243:114. 

\bibitem[Segal et al., 2004]{sbg}
Segal, M.R., Barbour, J. D. and Grant, R. M. (2004).
Relating HIV-1 sequence variation to
replication capacity via trees and forests.
{\emph{Stat Appl Genet Mol Biol.}} 3, 2.

\bibitem[Szendro et al., 2013]{ssf}
Szendro, I. G., Schenk, M. F., Franke, J.
Krug, J. and de Visser J. A. G. M. (2013).
Quantitative analyses of empirical fitness landscapes.
{\emph{J. Stat. Mech.}} P01005.


\bibitem[Weinreich et al., 2005] {wwc}
Weinreich, D. M., Watson R. A. and Chao, L. (2005).
Sign epistasis and genetic constraint on evolutionary 
trajectories. {\emph{Evolution}} 59, 1165--1174.


\bibitem[Weinreich et al., 2006]{wdd}
Weinreich, D. M., Delaney N. F., Depristo, M. A., and Hartl, D. L. (2006).
Darwinian evolution can follow only very few mutational paths to
fitter proteins. {\emph{Science 312}}: 111--114.


\bibitem[Wright, 1931]{w}
Wright, S. (1931). Evolution in Mendelian populations.
{\emph{Genetics}}, 16 97--159.


\bibitem[Ziegler, 1995]{z}
Ziegler G. (1995).
Lectures on Polytopes. Graduate Texts in Mathematics, 152, Berlin, New
York: Springer-Verlag.

\end{thebibliography}
\end{document}